\documentclass[12pt,a4paper]{article}

\usepackage[utf8]{inputenc}
\usepackage[T1]{fontenc}
\usepackage{amsmath}
\usepackage{float}
\usepackage{ae}
\usepackage{color}
\usepackage{graphicx}
\usepackage{bbm}

\usepackage[toc,page]{appendix}

\newcommand{\rd}{\ensuremath{\mathrm{d}}}

\newcommand{\hf}{\ensuremath{\hat{f}}}
\newcommand{\hh}{\ensuremath{\hat{h}}}
\newcommand{\hB}{\ensuremath{\hat{B}}}
\newcommand{\hb}{\ensuremath{\hat{b}}}
\newcommand{\hz}{\ensuremath{\hat{z}}}

\newcommand{\be}{\begin{equation}}
\newcommand{\ee}{\end{equation}}
\usepackage[colorlinks=true, linkcolor=blue, bookmarks=true]{hyperref}
\newcommand{\arXiv}[1]{\href{http://www.arXiv.org/abs/#1}{#1}}

\begin{document}

\begin{titlepage}
\vfill
\begin{center}
{\LARGE \bf Holographic thermalization in  a top-down confining model}

\vskip 20mm

{\large B.~Craps$\,^{a}$, E.~J.~Lindgren$\,^{a,b}$, A.~Taliotis$\,^{a}$}

\vskip 15mm

$^a$ Theoretische Natuurkunde, Vrije Universiteit Brussel, and \\ International Solvay Institutes, Pleinlaan 2, B-1050 Brussels, Belgium \\
$^b$ Physique Th\'eorique et Math\'ematique, Universit\'e Libre de Bruxelles, \\ Campus Plaine C.P.\ 231, B-1050 Bruxelles, Belgium \\

\vskip 15mm

{\small\noindent  {\tt Ben.Craps@vub.ac.be, Jonathan.Lindgren@vub.ac.be, Anastasios.Taliotis@vub.ac.be}}

\end{center}
\vfill

\begin{center}
{\bf ABSTRACT}
\vspace{3mm}
\end{center}
It is interesting to ask how a confinement scale affects the thermalization of strongly coupled gauge theories with gravity duals. We study this question for the AdS soliton model, which underlies top-down holographic models for Yang-Mills theory and QCD. Injecting energy via a homogeneous massless scalar source that is briefly turned on, our fully backreacted numerical analysis finds two regimes. Either a black brane forms, possibly after one or more bounces, after which the pressure components relax according to the lowest quasinormal mode. Or the scalar shell keeps scattering, in which case the pressure components oscillate and undergo modulation on time scales independent of the (small) shell amplitude. We show analytically that the scattering shell cannot relax to a homogeneous equilibrium state, and explain the modulation as due to a near-resonance between a normal mode frequency of the metric and the frequency with which the scalar shell oscillates.   

\end{titlepage}

\tableofcontents
\section{Introduction}

What happens to a strongly coupled field theory when it is brought far from equilibrium? This question is important in many areas of physics, including the formation of a quark gluon plasma in ultrarelativistic heavy ion collisions and quantum quenches in cold atom systems. It is a difficult question, however, because conventional techniques fail in the strongly coupled, far-from-equilibrium regime. In recent years, progress has been made using the gauge/gravity duality, also known as ``holography''. The simplest AdS/CFT models describe conformal field theories, and the original studies of holographic thermalization focused on those. Interestingly, when extrapolating to heavy ion collisions, one typically finds thermalization times that are short enough to be compatible with experiment  \cite{Kovchegov:2007pq, Albacete:2008vs, Chesler:2008hg, Beuf:2009cx, AbajoArrastia:2010yt,Albash:2010mv,Balasubramanian:2010ce,Heller:2011ju,Heller:2012km, Balasubramanian:2013rva}. Another noteworthy result is that short-
wavelength modes 
thermalize first in the simplest holographic models  \cite{Lin:2008rw, Balasubramanian:2010ce}. An obvious question is whether there are interesting new effects for non-conformal models, in particular for confining ones.

The study of holographic thermalization in confining models was initiated in \cite{Craps:2013iaa,Craps:2014eba}, where the hard wall model was considered, first in a weak field approximation \cite{Craps:2013iaa} and then using fully backreacted numerical simulations \cite{Craps:2014eba}. Following \cite{Bhattacharyya:2009uu} (see also \cite{Chesler:2008hg}), starting from the ground state, energy was injected by turning on a homogeneous scalar source of amplitude $\epsilon$ for a brief time interval $\delta t$. In the bulk, this leads to a planar shell falling towards the interior of AdS, which in the non-confining context of \cite{Bhattacharyya:2009uu} always led to the formation of a black brane, corresponding to thermalization in field theory. In the hard wall model, however, two regimes were found \cite{Craps:2013iaa,Craps:2014eba}. Certain shells collapsed into large black branes, while others kept scattering between the hard wall and the AdS boundary. In the scattering phase, for certain boundary 
conditions at the hard wall, the oscillating scalar expectation values underwent interesting modulation on time scales scaling like the inverse amplitude squared, due to resonant transfer of energy \cite{Craps:2014eba} similar to that discovered in \cite{Bizon:2011gg} for collapse in global AdS. 

The hard wall model is simple, but is sometimes criticized for being crude and ad hoc (an interior region of AdS being artificially cut away by the hard wall), and for being rather different from large-$N$ Yang-Mills theory in certain respects (see, for instance, Section~1.1 of \cite{Craps:2013iaa} for a brief discussion). We therefore decided to re-examine these issues in the context of the top-down AdS soliton model \cite{Witten:1998zw}, which underlies top-down holographic models for Yang-Mills theory \cite{Witten:1998zw} and QCD \cite{Sakai:2004cn}. In the bulk spacetime corresponding to the ground state, the radial direction and a circle combine into a cigar-shaped geometry, causing the radial direction to cap off smoothly at a radius that sets the confinement scale. While the starting point of the construction in \cite{Witten:1998zw} was the AdS$_7$ soliton (which after compactification on two circles left the radial direction and 3+1 large field theory dimensions), we will consider AdS solitons in 4, 
5, 6 and 7 dimensions.

Starting from the AdS soliton spacetime, we will again inject energy using a minimally coupled massless scalar field with amplitude of order $\epsilon$. An important difference with the hard wall model is that now the metric itself contains dynamics (in the sense that it is not completely determined by constraints), because there is no isotropy between the circle and the other spatial field theory dimensions. If and when a black brane forms, isotropy is restored in the metric components. Given a bulk solution, holographic renormalizaton can be used to extract field theory quantities such as the expectation values of the energy and of the pressure components.

We perform a fully backreacted numerical analysis, and identify a regime in which the infalling shell collapses into a black brane, possibly after one or more bounces, as well as a regime in which the infalling shell keeps scattering between the tip of the cigar (which we will henceforth refer to as the IR) and the AdS boundary. In the former case, we find that the pressure components relax to their (isotropic) equilibrium values according to their lowest quasinormal mode. In particular, the difference in pressure components along the noncompact and compact spatial dimensions of the AdS boundary relaxes to zero as an oscillating exponential. In the scattering phase, when the injected energy is sufficiently small, we show analytically that the shell cannot relax to a homogeneous equilibrium state, and we find numerically that the pressure anisotropy oscillates and undergoes modulation on a time scale that is $\epsilon$-independent in the limit of small amplitude $\epsilon$ and short injection time. This 
modulation time scale is very different from the $1/\epsilon^2$ time scale found for the  hard wall model, and indeed the physical mechanism is different as well: the oscillations are due to an almost resonance between the oscillation frequency of the scalar shell and the lowest normal mode frequency of the dynamical metric component. Just above the threshhold for black hole formation, we also find solutions that bounce a few times against the AdS boundary before collapsing into a black brane, similar to solutions found in global AdS \cite{Bizon:2011gg}.

Recently, several other papers have studied holographic thermalization in non-conformal models. Based on a quasinormal mode analysis of top-down non-conformal (but gapless) models, it was conjectured in \cite{Buchel:2015saa} that, as soon as a horizon is formed in the bulk, deviations from conformality do not significantly affect thermalization times. Similarly, the numerical analysis in \cite{Fuini:2015hba} found that the equilibration dynamics of $N=4$ SYM theory does not change much when chemical potentials or magnetic fields are added. In \cite{Janik:2015waa}, quasinormal modes were computed for bottom-up models mimicking the equation of state of QCD, and a non-trivial dependence on the equation of state was found. A confining bottom-up model for QCD was studied nonlinearly in \cite{Ishii:2015gia}, but only for initial conditions that already contain a small black hole; in this case, good agreement with a quasinormal mode analysis was found. In the present paper, we study top-down confining models at the 
nonlinear level, in regimes with and without horizons. If and when a horizon forms, the subsequent dynamics is well-described by a quasinormal mode analysis, and the confinement scale does not play much of a role. In the parameter regime in which no horizon forms, the dynamics is dramatically different.

The remainder of this paper is organized as follows. In Section~\ref{setup} we  describe our setup, including our metric ansatz, the equations of motion and how to extract field theory quantities from the bulk solutions. In Section~\ref{numerics} we briefly discuss the numerical methods we have used. Section~\ref{phasediagram} starts introducing the reader to our numerical results and to the two different phases. In Section~\ref{blackholephase} we discuss the black hole phase and in Section~\ref{scatteringphase} the scattering phase. We conclude in Section~\ref{conclusions}. A number of technical details are contained in three appendices.

\section{Holographic setup}\label{setup}
The model we will consider is the Einstein-Hilbert action with a minimally coupled massless scalar field,
\begin{equation}
S=\frac{1}{2\kappa^2}\int \rd^{d+1}x\sqrt{-g}\left(R-2\Lambda-\frac{1}{2}(\partial \phi^2)\right)-\frac{1}{\kappa^2}\int_{\partial}\rd^dx\sqrt{-\gamma}\mathcal{K},\label{action}
\end{equation}
where the cosmological constant $\Lambda$ is related to the AdS radius $L$ by $\Lambda=-d(d-1)/2L^2$. 
The boundary term is the Gibbons-Hawking-York term, which is necessary to render the variational principle well defined \cite{Gibbons:1976ue}, but it will not play a role in this work. We will start with the AdS soliton as an initial condition, and then inject energy into the system by perturbing the scalar field.

The AdS soliton background corresponds to a Euclidean black brane with an extra time direction. An explicit metric can be written as
\begin{equation}
\rd s^2=\frac{L^2}{z^2}\left(-\rd t^2+\frac{\rd z^2}{1-\frac{z^d}{z_0^d}}+(1-\frac{z^d}{z_0^d})\rd \theta^2+\rd \vec{x}_{d-2}^2\right).\label{solitonmetric}
\end{equation}
We will henceforth work in units with $L=1$. The AdS boundary is located at $z=0$ and the confinement scale is set by $z_0$ (which would correspond to the horizon if we Wick rotated back the above to a black brane). Note that the $\theta$ coordinate is compact in order to avoid a conical singularity, and this metric breaks rotational invariance between the $\vec{x}$ coordinates and the $\theta$ coordinate. This will have the implication that in order to solve for the time-dependence of the metric, we will need the second order dynamical Einstein equations. This should be contrasted with rotationally invariant metrics, for which the metric can be determined using first order constraint equations alone.

The massless scalar field will be dual to a marginal operator in the dual field theory. To quench the system we will use a source $J$ coupled to this operator and turn it on for a short period of time. While we imagine the source to vanish outside a finite time interval, in our numerical computations we choose for simplicity a Gaussian profile of the form
\begin{equation}
J(t)=\epsilon e^{-\frac{t^2}{\delta t^2}},\label{source}
\end{equation}
which is indeed negligibly for $|t|\gg \delta t$.
The total injected energy will scale like $E\sim \epsilon^2/\delta t^d$ for small $\epsilon$ and small $\delta t$ \cite{Bhattacharyya:2009uu}. In the dual gravitational description, this source term corresponds to the value of $\phi$ at the AdS boundary. After the source has been turned off, the system's energy will have increased and the gravitational bulk solution will have a nontrivial time dependence, governed by the action \eqref{action}. The main question is how this time-dependent solution behaves, in particular if it collapses into a black brane solution or not. To avoid an extra scalar field, we will also consider quenching the metric, and compare this to the case of perturbing the scalar. We can inject energy by turning on a short time dependence for $\eta_{\theta\theta}/\eta_{x_jx_j}$, where $\eta$ is the boundary metric, which breaks the isotropy of the boundary metric between the $\theta$ and $\vec{x}$ coordinates (see Section~\ref{sec:ansatz}, in particular \eqref{bsource}). This can also be 
interpreted as quenching the size of the compactified dimension. In this case, only the gravitational mode will be turned on, and the dynamics will be qualitatively different from the case where both the scalar and the metric mode are turned on. Although we will just very briefly consider such quenches in the metric, it is important to remember that it is possible in this setup to inject energy via the metric without a scalar field and without breaking additional rotational symmetries.

\subsection{Ansatz and equations of motion}\label{sec:ansatz}
To solve the Einstein equations we need to choose specific coordinates. We can constrain the form of the metric by using the symmetries of the problem and by suitable gauge transformations (diffeomorphisms), but otherwise the metric will be completely general. In particular, our metric will only depend on time and on the radial bulk coordinate. Also, note in particular that due to parity invariance in the $\theta$ and $\vec{x}$ coordinates, we can set all off-diagonal terms involving these coordinates to zero. We may then use our gauge transformation to bring the metric to a diagonal form with three free functions. Note that the absence of rotational symmetry between the $\theta$ and $\vec{x}$ coordinates in the AdS soliton background forces us to choose a more general ansatz than in setups with rotational symmetry in all spatial coordinates \cite{ Bhattacharyya:2009uu, Wu:2012rib,  Craps:2013iaa, Craps:2014eba, Ishii:2015gia}, in which case there are usually only two free functions in the metric, which are 
completely determined by the constraint equations (there are no propagating degrees of freedom in the metric). We have found that the following ansatz is useful:
\begin{equation}
\rd s^2=\frac{L^2}{z^2}\left(-h(z,t)^2\rd t^2+ \frac{f(z,t)^2}{1-\frac{z^d}{z_0^d}}\rd z^2+ (1-\frac{z^d}{z_0^d})e^{(d-2)b(z,t)}\rd \theta^2+e^{-b(z,t)}\rd \vec{x}_{d-2}^2\right),\label{zansatz}
\end{equation}
with the initial conditions that $f=1$, $h=1$ and $b=0$ before the injection. We will refer to the boundary at $z=0$ as the UV and the point $z=z_0$ as the IR. The coordinate $\theta$ is periodic with period $4\pi z_0/d$ to avoid a conical singularity.  This form of the metric has a remaining gauge symmetry corresponding to rescaling of all coordinates. In the numerics, we will use this to set $z_0=1$, but for now we will keep $z_0$ explicit. The coordinates in (\ref{zansatz}) are very badly behaved at $z=z_0$, however, so for numerics we will use a different ansatz, see Section~\ref{numerics} and Appendix~\ref{reom}. 

To inject energy via the metric, we do this via the function $b(z,t)$, namely we can assume a boundary profile on the form
\begin{equation}
b(0,t)=\epsilon e^{-\frac{t^2}{\delta t^2}},\label{bsource}
\end{equation}

and turn off the scalar. It turns out to be convenient to define the following variables
\begin{equation}
\begin{array}{ccc}\label{PPi}
 \Pi=\dot{\phi}\frac{f}{h},  &\hspace{20pt} P=\dot{b}\frac{f}{h},\\
\Phi=\phi', &\hspace{20pt} B=b',\\
\end{array}
\end{equation}
where  $'$ means derivative with respect to the $z$ coordinate.\footnote{We warn the reader that $'$ will have a different meaning in Section~\ref{static} and Appendix~\ref{reom}.}  
Introducing
\begin{equation}
G(z)=1-\frac{z^d}{z_0^d},
\end{equation}
the equations of motion following from the ansatz \eqref{zansatz} are
\begin{align}
\dot{f}=&z^{1-2d}\frac{(d-2)(d-1)Gh}{2(G z^{-2(d-1)})'}(PBz+2P)+\frac{z^{2-2d}\Phi\Pi Gh}{(G z^{-2(d-1)})'}+\frac{d-2}{2}Ph\label{zdotf},\\\nonumber
\frac{h'}{h}=&\frac{1}{z^{2(d-1)}(Gz^{-2(d-1)})'}\left(\frac{(d-1)(d-2)}{2}(P^2+GB^2+\frac{4GB}{z})+G\Phi^2+\Pi^2\right)+\\
&+(d-2)B-\frac{f'}{f},\\
\frac{h'}{h}=&\frac{2d(d-1)(f^2-1)}{z^{2d}(Gz^{-2(d-1)})'}+\frac{f'}{f},\\
\dot{P}=&\frac{1}{d-1}\left(\frac{(Ge^{(d-1)b})'he^{-(d-1)b}}{fz^{d-1}}\right)'z^{d-1},\\
\dot{\Pi}=&\left(\frac{hG\Phi}{fz^{d-1}}\right)'z^{d-1},\\
\dot{B}=&\left(\frac{Ph}{f}\right)',\\
\dot{\Phi}=&\left(\frac{\Pi h}{f}\right)'.
\end{align}
Note that in this gauge, only derivatives of $b$ appear in the equations of motion, so we do not need to integrate at every time step to obtain $b$. This is the reason for the particular parametrization in \eqref{zansatz}, which makes the equations of motion decouple nicely. A similar ansatz is used in \cite{Chesler:2008hg}.

Evaluating equation \eqref{zdotf} at the point $z=z_0$, and using (\ref{PPi}), we obtain
\begin{equation}
\dot{f}_{z=z_0}=(\frac{d-2}{2}P h)_{z=z_0}=(\frac{d-2}{2}\dot{b}f)_{z=z_0},
\end{equation}
so that
\begin{equation} \label{fB}
f_{z=z_0}=C {e^{\frac{d-2}{2}b}}{\Large |}_{r=0}
\end{equation}
for some constant $C$. Since initially $f=1$ and $b=0$, we have that $C=1$. Thus we can state this result as
\begin{equation}
\left(fe^{-\frac{d-2}{2}b}\right)_{z=z_0}=1,\label{feB2}
\end{equation}
which will be crucial for the analysis in Section~\ref{static}. This condition actually is the statement that the regularity (absence of a conical singularity) at $z=z_0$ is preserved in time. This can be easily seen in the ansatz \eqref{ransatz}, which is used in the numerical analysis, and we refer the reader to Section~\ref{numerics} for further discussion.
 
\subsection{Boundary expansion and holographic renormalization}\label{boundaryexp}
To compute field theory observables, one resorts to a process called ``holographic renormalization'' \cite{deHaro:2000xn, Skenderis:2002wp}, which requires adding counterterms to the action to cancel divergences from the near-boundary region. These counterterms, which in odd dimensions give finite contributions to the various one-point functions, must be evaluated explicitly for every dimension and quickly become quite involved for increasing dimension. In addition, these contributions make the one-point functions scheme-dependent. However, when the source is turned off, the first non-trivial term in the boundary expansion is of order $z^d$ and no counterterms are needed. Since we will be interested in the evolution of the one-point functions after the source has been turned off, we will therefore be able to ignore the counterterms. In even dimensions the counterterms do not give finite contributions to the one-point functions even when the source is nonzero and thus in this case the counterterms can always 
be ignored \cite{deHaro:2000xn, Papadimitriou:2011qb}. For $d=3$ we provide the full asymptotic boundary expansion in Appendix \ref{asympt3}, which will be used in some of the figures.

The asymptotic behaviour of the various fields after the source has been turned off is given by
\begin{align}
f=&1+\frac{E}{2(d-1)}z^d+\ldots,\label{fUV}\\
h=&1-\frac{E}{2(d-1)}z^d+\ldots,\\
b=&b_dz^d+\ldots,\\
\phi=&\phi_dz^d+\ldots,
\end{align}
where the $z^d$ coefficients of $f$ and of $h$ have been related by the equations of motion. We will see later that $E$ will be the total injected energy, while the coefficient $\phi_d$ is related to the vacuum expectation value of the dual operator. We also have that $\dot{E}=0$, which follows from the equations of motion or from holographic Ward identities.

To identify the stress energy components at the boundary, we want to write the metric in the Fefferman-Graham gauge
\begin{equation}
\rd s^2=\frac{\rd \zeta^2}{\zeta^2}+\frac{1}{\zeta^2}g_{\alpha\beta}\rd x^\alpha \rd x^\beta.
\end{equation}
Doing this asymptotically, we can identify $z=\zeta-\zeta^{d+1}\frac{1}{2d}(\frac{E}{d-1}+\frac{1}{z_0^d})+O(\zeta^{d+2})$, which gives us the metric
\begin{align}
\rd s^2=\frac{\rd \zeta^2}{\zeta^2}-\frac{1}{\zeta^2}{\big\{}& \left(1-(E-\frac{1}{z_0^d})\frac{\zeta^d}{d}+O(\zeta^{d+1})\right)\rd t^2\nonumber\\&+\bigg(1+(\frac{E}{d-1}+\frac{1-d}{z_0^d})\frac{\zeta^d}{d}+(d-2)b_d\zeta^d+O(\zeta^{d+1})\bigg)\rd \theta^2\nonumber\\&+\bigg(1+(\frac{E}{d-1}+\frac{1}{z_0^d})\frac{\zeta^d}{d}-b_d\zeta^d+O(\zeta^{d+1})\bigg)\rd \vec{x}^2 {\big\}}. \label{asymptmetric}
\end{align}
Now it is easy to read off the non-zero stress energy components of the boundary field theory \cite{deHaro:2000xn, Papadimitriou:2011qb}:
\begin{subequations}\label{Ttt}
\begin{align}
\langle T_{tt}\rangle=& \frac{1}{16\pi G_N}(E-\frac{1}{z_0^d}),\\
\langle T_{\theta\theta}\rangle=&\frac{1}{16\pi G_N}(\frac{E}{d-1}+\frac{1-d}{z_0^d}+d(d-2)b_d),\\
\langle T_{xx}\rangle=& \frac{1}{16\pi G_N}(\frac{E}{d-1}+\frac{1}{z_0^d}-db_d),
\end{align}
\end{subequations}
from which we see that $E$ is indeed the total injected energy (up to a factor of $1/16\pi G_N$), and $-\frac{1}{16\pi G_N}\frac{1}{z_0^d}$ is the  initial AdS soliton energy density. Note also that $\langle T_\mu^\mu \rangle=0$.\footnote{This is not generally true while the source is turned on; see, for example, Eqs (20)-(23) of \cite{Craps:2014eba}.} The vacuum expectation value of the scalar is
\begin{equation}\label{vev}
\langle \mathcal{O} \rangle=\frac{1}{16\pi G_N}\phi_d.
\end{equation}
Note that taking the difference $\langle T_{\theta\theta}\rangle-\langle T_{xx}\rangle$ cancels the total injected energy $E$ and isolates the dynamical mode $b$, which is why we will prefer to plot this quantity instead of the individual pressure components.

\subsubsection{Temperature of black brane solutions}
As seen in \eqref{Ttt}, the energy density will be positive for energies $E>1/z_0^d$, and we expect that black branes will form. A black brane can be written as the metric
\begin{equation}
\rd s^2=\frac{1}{\xi^2}\left(-\rd t^2(1-\frac{\xi^d}{\xi_h^d})+\frac{\rd \xi^2}{1-\frac{\xi^d}{\xi_h^d}}+\rd\theta^2+\rd \vec{x}_{d-2}^2\right).\label{bhmetric}
\end{equation}
Note in particular that in the case of dynamically evolving from the AdS soliton background into the black brane \eqref{bhmetric}, isotropy between the $\vec{x}$ and $\theta$ coordinates must then be restored. From \eqref{Ttt} this means that we must have $b_d=\frac{1}{(d-1)z_0^d}$ and this is indeed verified numerically. The temperature of such a black brane, as obtained by the standard procedure of requiring the absence of a conical singularity for the Euclidean version of \eqref{bhmetric}, is given by $T=d/4\pi\xi_h$. Asymptotically, the radial coordinates $\xi$ and $\zeta$ are related by $\xi=\zeta-\zeta^{d+1}/2d\xi_h^d$, from which, comparing with \eqref{asymptmetric}, we can obtain the temperature of the black brane,
\begin{equation}
T=\frac{d}{4\pi \xi_h}=\frac{d}{4\pi}\left[\frac{E-\frac{1}{z_0^d}}{d-1}\right]^{1/d}.
\end{equation}


\section{Numerical methods}\label{numerics}
In this section we will list some important tricks that we had to employ to achieve stable numerical evolution. We used a fourth order finite difference method to discretize the radial direction, and then we used the ordinary differential equation solver \verb!scipy.integrate.ode! from the \verb!Python! library \verb!scipy!  
\cite{scipy} to evolve the resulting system of ordinary differential equations in time. We have as initial conditions $f=1$, $h=1$, $b=0$ and $\phi=0$, corresponding to the AdS soliton geometry. The boundary conditions we impose in the UV are $f(0,t)=1$ and $h(0,t)=1$ as well as $\phi(0,t)=J(t)$ and $b(0,t)=0$ ($b(0,t)=J(t)$ and $\phi\equiv0$ if we quench the metric instead of the scalar), and the source is always taken as a gaussian $J(t)=\epsilon e^{-t^2/\delta t^2}$. In the IR, we do not have to impose any boundary conditions, since regularity already follows from the equations of motion. However, there are some potential sources for numerical instability and inaccuracy, the coordinate singularity in the IR and the AdS boundary being two examples.

\subsection{The coordinate singularity}
The $z$ coordinate ansatz \eqref{zansatz} is very inconvenient in the IR. The reason is that at this point the geometry looks locally like Minkowski space in cylinder coordinates, with rotational invariance in the $(z,\theta)$ plane. However, the relation to the radial coordinate in this locally flat space is $z=z_0(1-r^2)$, and thus $\rd z=-2rz_0 \rd r$. This means that a small grid spacing in $z$ will be mapped to a very large grid spacing in $r$ (which is the natural coordinate around the point $z=z_0$), so a linearly spaced discretization in the $z$ coordinate will become incredibly bad at this point. We thus found it convenient to instead work with the coordinate $r=\sqrt{1-z/z_0}$, and use the metric ansatz
\begin{equation}\label{ransatz}
\rd s^2=\frac{1}{s(r)^2}(-h(r,t)^2\rd t^2+\frac{4f(r,t)^2}{dg(r)} \rd r^2+ r^2g(r)e^{(d-2)b(r,t)}\rd \tilde{\theta}^2+e^{-b(r,t)}\rd \vec{x}_{d-2}^2),
\end{equation}
where $s(r)=z_0(1-r^2)$ and $g(r)=(1-(1-r^2)^d)/(z_0^2r^2d)$. The advantage of this parametrization of the metric is that now $g(r)$ is a finite slowly varying non-zero function and $g(0)=1/z_0^2$. The new periodic coordinate $\tilde{\theta}$ now has period $4\pi z_0^2/d^{3/2}$. While the coordinate system \eqref{zansatz} is convenient to derive analytic results and to extract the boundary field theory observables, the coordinate system \eqref{ransatz} will be used for the numerical evolution. It is also clear now in these coordinates, that the regularity condition (absence of conical singularity), means that $fe^{(d-2)b/2}$ remains constant in time, which is exactly the statement \eqref{feB2}. The equations of motion for the ansatz \eqref{ransatz} can be found in Appendix \ref{reom}. The functions $f$, $P$, $\Pi$, $\Phi$ and $B$ are evolved in time, while the function $h$ is solved for at each time step using equation \eqref{heq}, and equation \eqref{constraint} is checked for consistency during the time 
evolution.

There is also a convenient trick that can be employed to compute derivatives close to an origin of a polar coordinate grid. Usually if one were to employ finite differences close to a boundary, one would have to resort to non-symmetric stencils which can induce instabilities or numerical inaccuracies. However, at $r=0$ we do not have a boundary, and we can imagine continuing $r$ past $r=0$ to negative values and thus it is possible to still use central difference schemes when computing derivatives close to $r=0$. An equivalent way of reaching the same result is to use the fact that all functions must be even functions of $r$ when computing the derivatives.

\subsection{Radial discretization}
We have found that high order finite difference discretization has worked well. However, to avoid high frequency spurious oscillations, we have found that it is convenient to put different functions on two different grids. To motivate this, consider first a function $v(t,z)$ satisfying the free wave equation $\ddot{v}=v''$. Defining $V=v'$ and $P=\dot{v}$ we obtain
\begin{align}
\dot{V}=P',\hspace{20pt}\dot{P}=V',
\end{align}
which should be compared to the equations of motion for the scalar field and the metric component $b$. Now, if we discretize the $z$ coordinate by $\{z_j\}_{j=0}^n$, and consider the derivative approximation $(z_{j+1}-z_{j})/\Delta z$, this will compute an approximation to the derivative at the point $(z_j+z_{j+1})/2$. We thus see that it might be convenient to put $V$ and $P$ on two different grids, one on $\{z_j\}_{j=0}^n$, and one one $\{x_j\}_{j=0}^n$ where $x_j=(z_j+z_{j+1})/2$, to improve the accuracy of the derivative approximations. If one were to use a central difference scheme, we find that it typically induces high frequency noise. This high frequency noise is still present when using higher order central difference schemes, but disappears when putting $V$ and $P$ on different grids (also when we use a higher order finite difference scheme).

In our more complicated setup, the same reasoning holds for the free wave equation in AdS, and we have found it very useful to employ the same trick even when including backreaction. Thus we have put $\Phi$ and $B$ on one grid, and $\Pi$, $P$ and $f$ on the other. Function values are then interpolated to the other grid when necessary. This proves to result in very stable evolution and the high frequency noise that is present when using central difference schemes with all functions on the same grid disappears.

 \subsection{Extracting boundary data}
 To extract the boundary data, we will have to compute quantities like $(f(z)-f(0))/z^d$ when $z\rightarrow0$. This becomes increasingly difficult when the dimension increases, since we are taking the ratio of two very small numbers. In particular for $d=6$, there is high frequency noise which makes it difficult to extract the observables. For the simulations of black hole formation (Fig.~\ref{qnm}), we therefore found it appropriate to use a Savitzky-Golay \cite{SG} filter to get rid of this noise and to make the boundary observables more smooth in time.

\section{Phase diagram}\label{phasediagram}
\begin{figure}[H]
\centering
\includegraphics[scale=0.3]{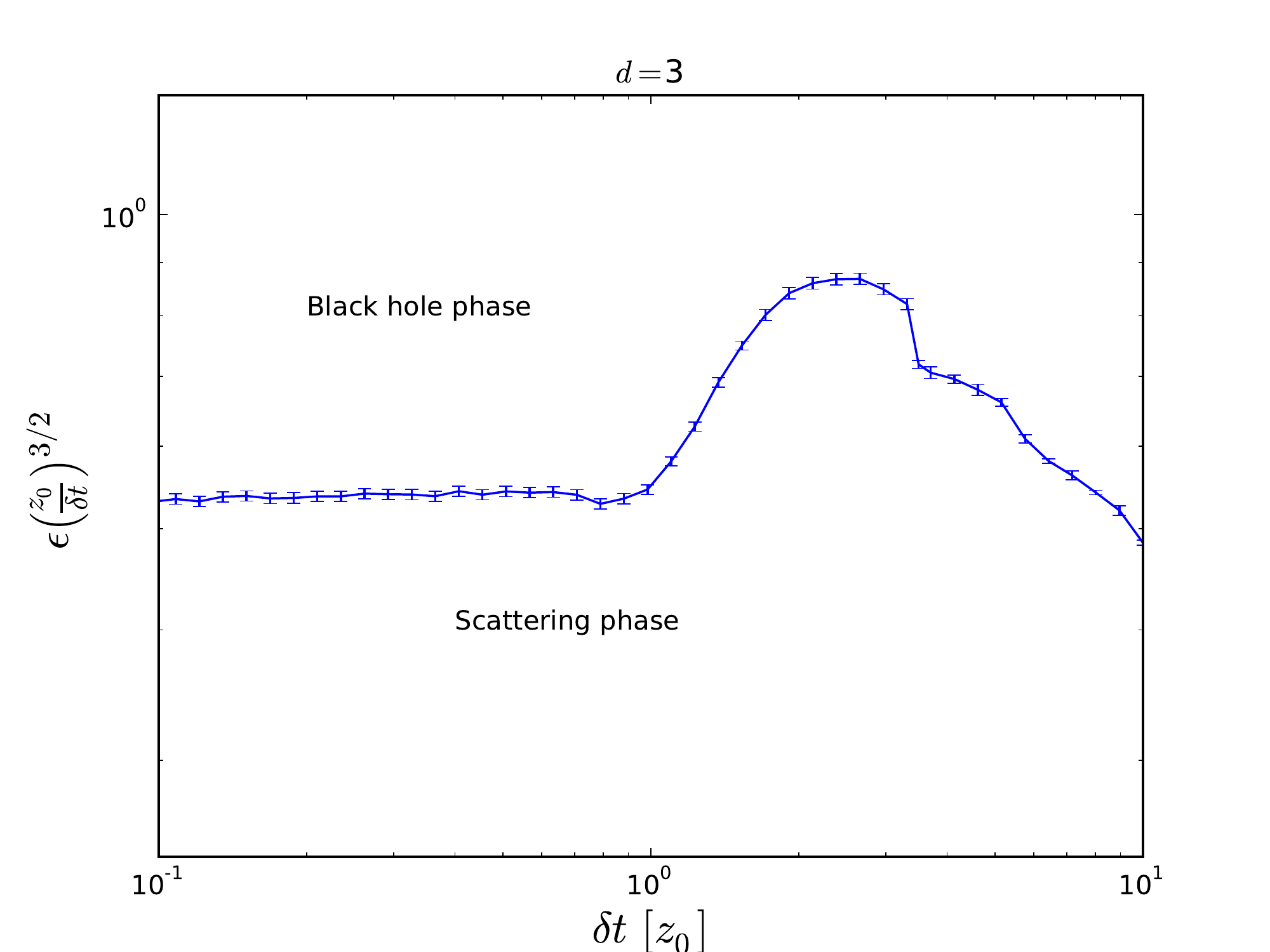}
\includegraphics[scale=0.3]{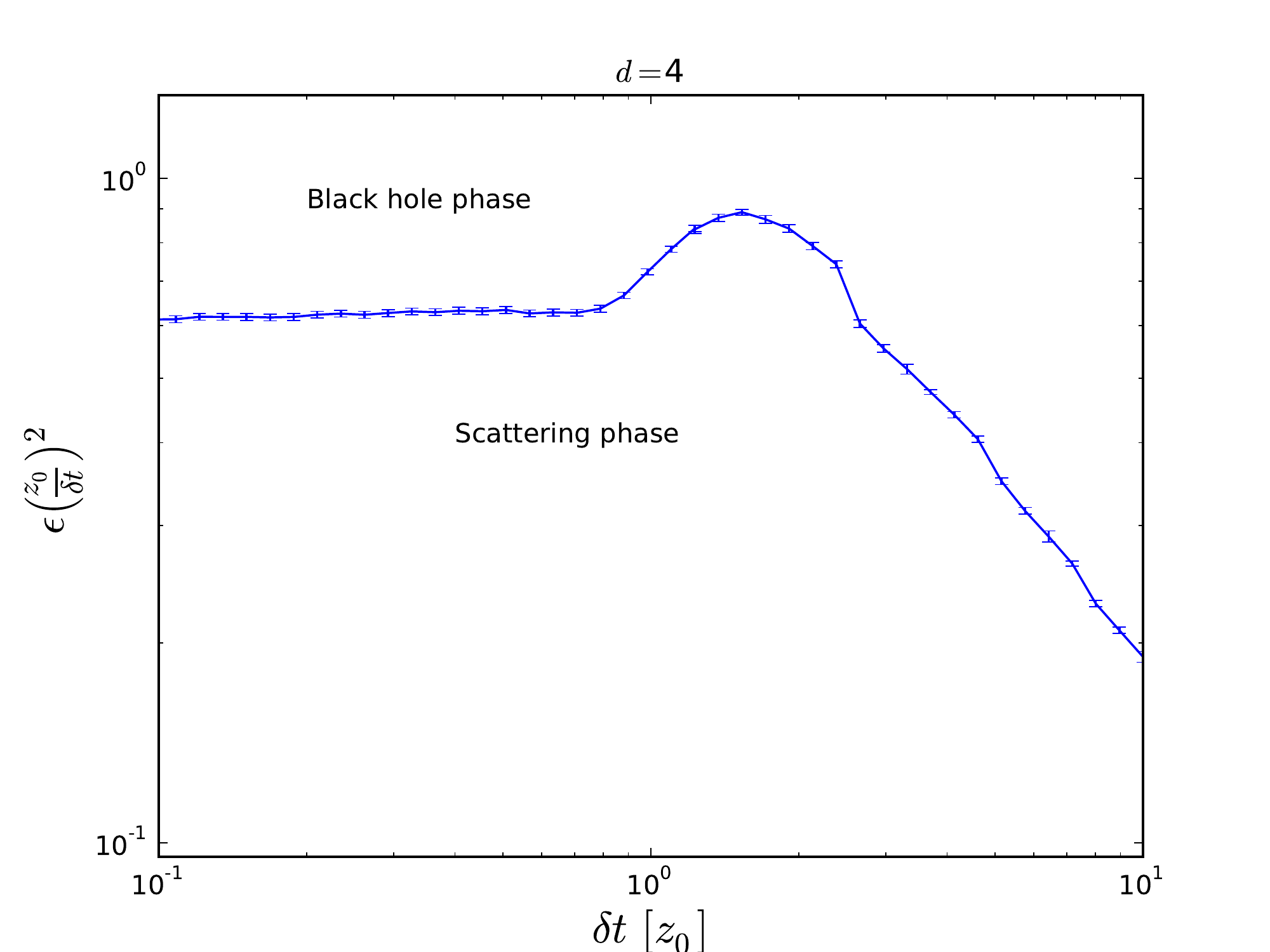}
\includegraphics[scale=0.3]{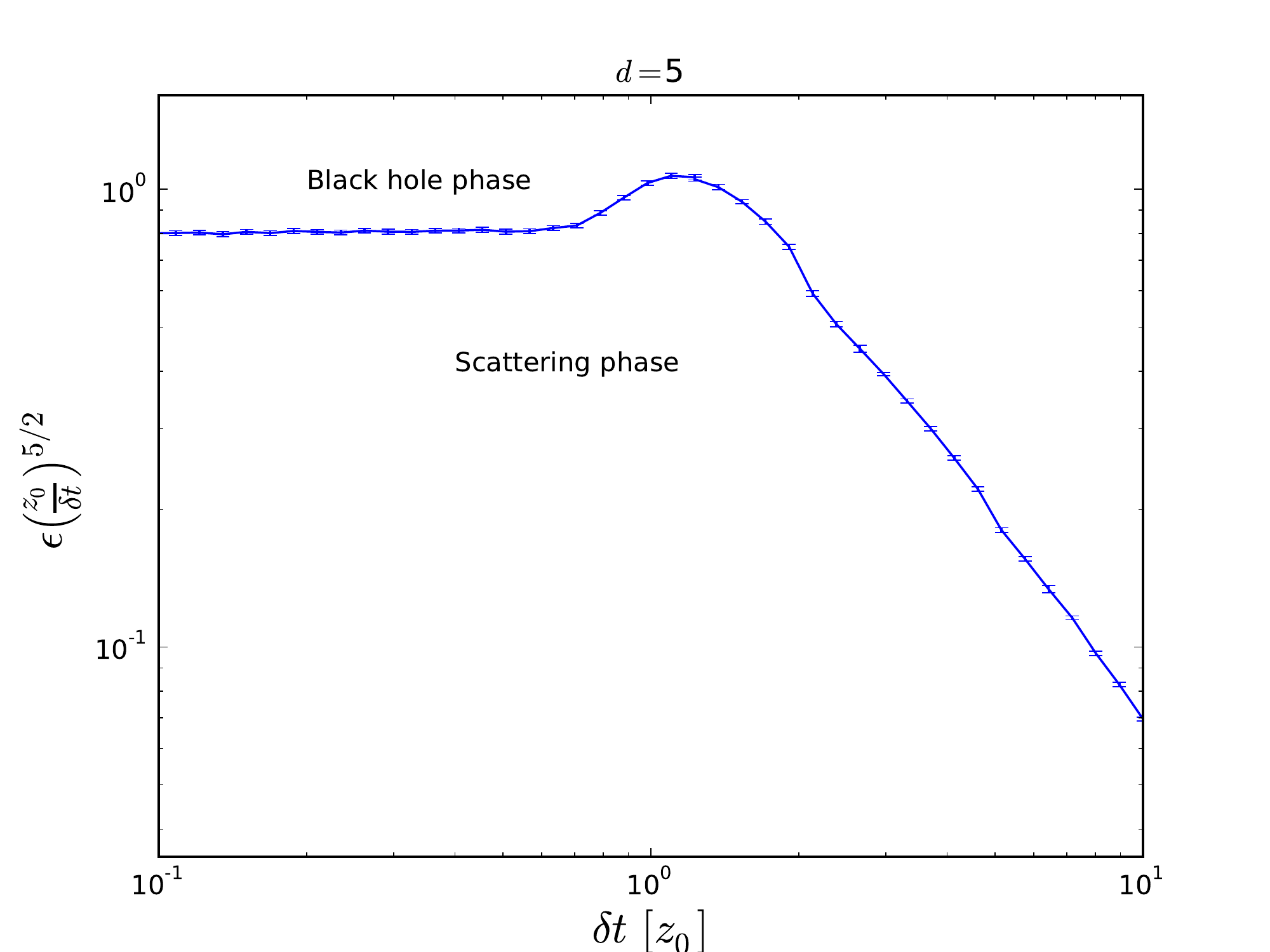}
\includegraphics[scale=0.3]{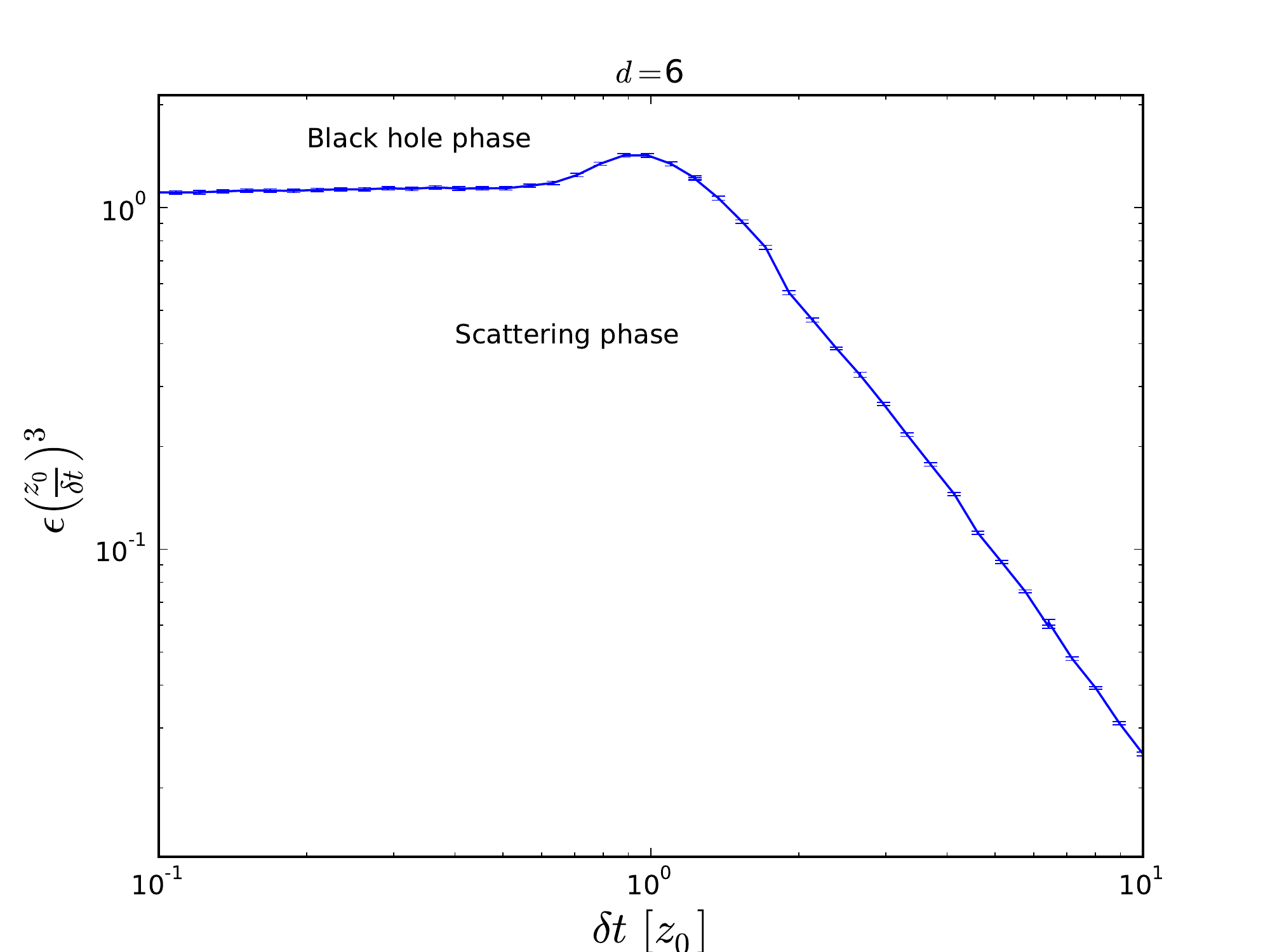}
\caption{The separation between the black hole phase and the scattering phase. For small $\delta t$, we see that $\epsilon\sim\delta t^{d/2}$, which is expected since the total injected energy goes like $E\sim\epsilon^2/\delta t^d$  \cite{Bhattacharyya:2009uu}. For large $\delta t$, we find the relation $\epsilon\sim\delta t$. }
\label{phase_diag} 
\end{figure}

When injecting energy into (the Poincar\'e patch of) vacuum AdS, we always form a black brane. However, since the energy density of the AdS soliton is negative, and any black brane has positive energy density, there should be a threshold for black hole formation when injecting energy into the AdS soliton. The obvious question is then, what solution do we obtain below the threshold? In the probe limit, the scalar field will just bounce forever between the boundary and the IR, so one could ask if this behaviour will still remain when turning on backreaction, or if the system will equilibrate into some static solution after a long time. In Section~\ref{static}, we prove that the system cannot equilibrate into any static solution. We will thus refer to these solutions as the ``scattering phase'', and the solutions that thermalize into black holes as the ``black hole phase''. In Fig.~\ref{phase_diag}, we show the separation between the two different phases, in terms of the  parameters $\epsilon$ and $\delta t$. 
For small $\delta t$ we have the relation $\epsilon\sim\delta t^{d/2}$, which is expected since the injected energy (which is the only parameter associated to the shell in the thin shell limit) goes like $E\sim\epsilon^2/\delta t^d$ \cite{Bhattacharyya:2009uu}. The shapes of the phase diagrams resemble those found for the hard wall model in \cite{Craps:2014eba}. In particular, for large $\delta t$ we find numerically the relation $\epsilon\sim\delta t$, which is the same as in the hard wall model with Neumann boundary conditions.

\begin{figure}[H]
\centering
\includegraphics[scale=0.7]{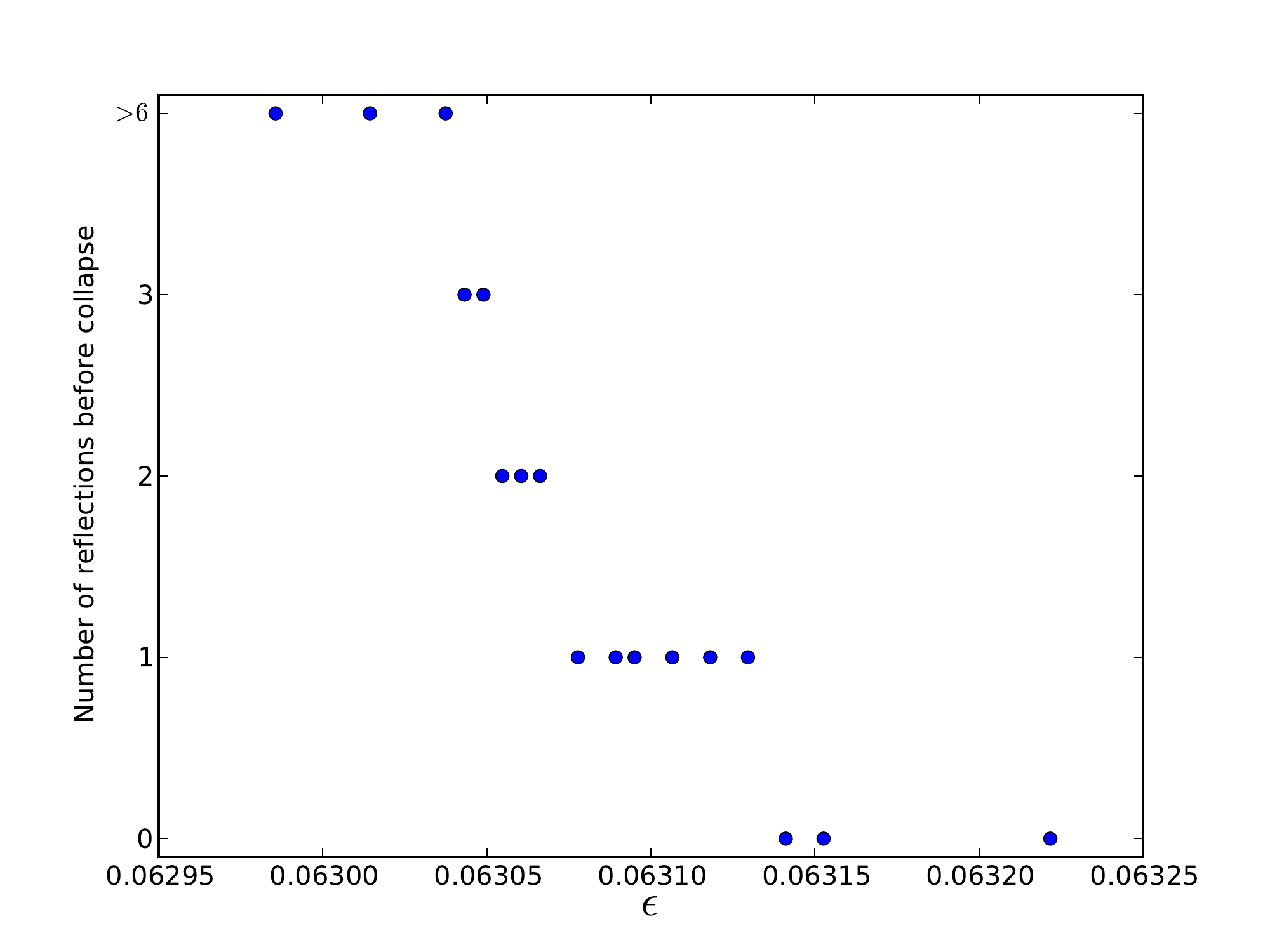}
\caption{Number of reflections at the boundary of the scalar field wave packet before black brane formation, as a function of $\epsilon$ for $d=3$ and $\delta t=0.24 z_0$. Note that for $\epsilon$ smaller than $0.06304$, there is a parameter region where the injected energy density is above the black brane threshold, but nevertheless the solutions seem to scatter for as long as we have been able to follow them. This region is relatively large, since the threshhold where the energy density becomes negative is $\epsilon\approx0.0607$.}
\label{nrofbounces} 
\end{figure}

Another interesting question is if we can find scattering solutions above the energy threshold. Intuitively, right above the threshhold, a wave packet should bounce before collapsing into a black brane due to the finite width of the wave packet, and this is indeed what we find: Right above the threshhold when black brane formation is possible (the energy density is positive), there is a region where solutions reflect many times against the boundary without collapsing (although we are not able to say whether they eventually collapse, due to numerical difficulties in following the solutions for a long time). We have also found solutions that bounce a few times and then collapse into a black hole, similar to what was found in global AdS \cite{Bizon:2011gg}. In Fig.~\ref{nrofbounces} we plot the number of scatterings before collapse, as a function of amplitude $\epsilon$ for fixed $\delta t=0.24 z_0$. We see that when decreasing $\epsilon$ the number of reflections against the boundary before collapse varies 
between 0 and 3, and then for smaller $\epsilon$ there is a large region where the solutions do not seem to collapse.

\begin{figure}[H]
\centering
\includegraphics[scale=0.7]{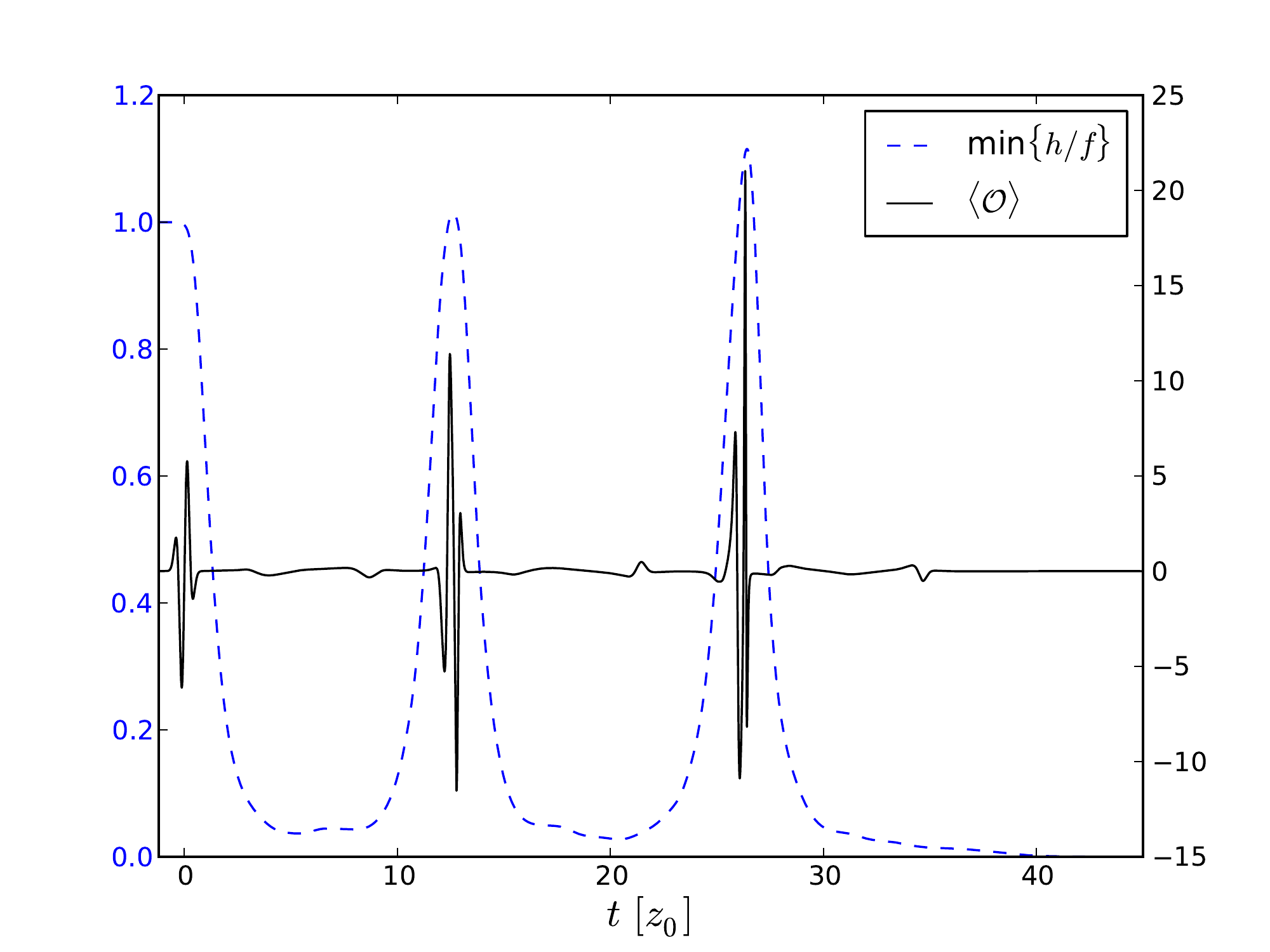}
\caption{Example of a quench where the scalar wave packet reflects twice at the boundary before collapsing into a black brane. Time is here in units of $z_0$. The parameters are $d=3$, $\epsilon=0.06305472$ and $\delta t=0.24 z_0$. The left axis is for $h/f$ and the right axis is for $\langle \mathcal{O} \rangle$ in units of $1/(16\pi G_N z_0^d)$. Vanishing of $h/f$ signals the formation of an apparent horizon.}
\label{reflections} 
\end{figure}

In Fig.~\ref{reflections}, we show  the vacuum expectation value of the scalar operator and min$\{f/h\}$ as a function of time, for a solution that bounces twice before collapsing into a black hole. 
After two reflections (identified by the sharp peaks in the vacuum expectation value) we see that min$\{f/h\}$ approaches zero, which indicates the formation of an apparent horizon. If the wave packet is very close to collapsing to a black hole while it scatters in the IR, the wave packet usually becomes very squeezed and comes out almost like a shock wave, resulting in the very sharp peaks in the expectation value $\langle \mathcal{O} \rangle$. 

\section{Black hole phase}\label{blackholephase}

\begin{figure}[H]
\centering
\includegraphics[scale=0.7]{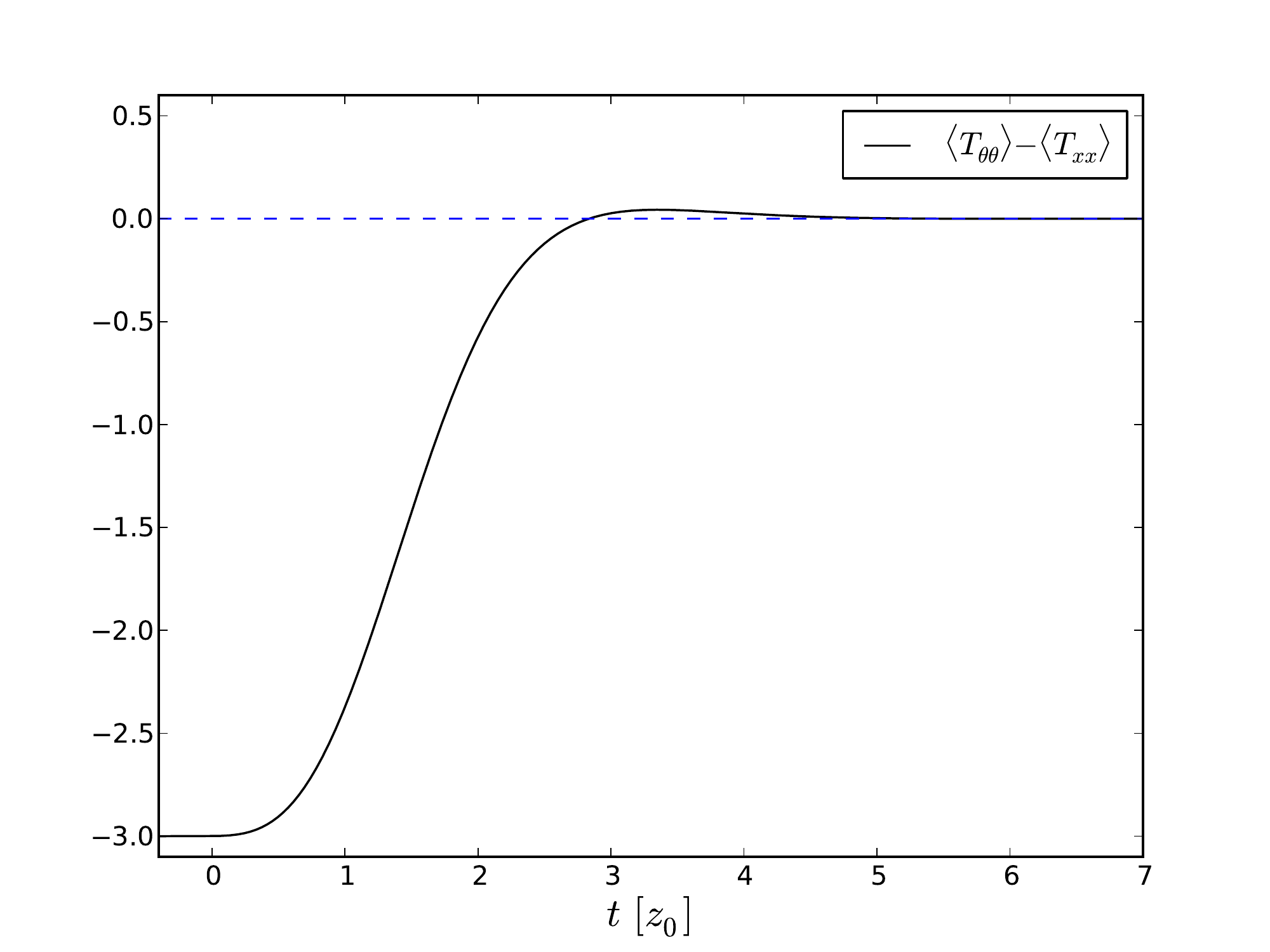}
\caption{Example of a quench ($\delta t=0.1 z_0$, $\epsilon=0.02$) resulting in black hole formation for $d=3$. Time is here in units of $z_0$, and the vacuum expectation values are in units of $1/(16\pi G_N z_0^d)$. The temperature of the black brane in this example is $T\approx0.15/z_0$. }
\label{blackhole} 
\end{figure}

In the black hole phase, the space-time will collapse into a black brane, and a horizon will form. The resulting solution will be an AdS$_{d+1}$ black brane. This in particular means that isotropy between the $\theta$ coordinate and the $\vec{x}$ coordinates will be restored, which in particular means from equation \eqref{Ttt} that $b_d=\frac{1}{(d-1)z_0^d}$, and this is indeed verified numerically. Thus the pressure anisotropy $\langle T_{\theta\theta}-T_{xx}\rangle$ will dynamically evolve from $-d/16z_0^d\pi G_N$ to 0. A relevant question is what this isotropization process looks like. In Fig.~\ref{blackhole}, we show a typical evolution of the pressure anisotropy for $d=3$. We see that the system quite rapidly enters a regime where it is isotropic up to some small fluctuations. In Section~\ref{sec:quasinormal}, we will compare these small fluctuations with the quasinormal modes of the resulting black brane. Our numerics will not allow us to follow the evolution for very long times after a black hole has 
formed, 
but long enough to see the quasinormal mode behaviour.

\subsection{Quasinormal modes and late time behaviour}\label{sec:quasinormal}
\begin{figure}[H]
\centering
\includegraphics[scale=0.33]{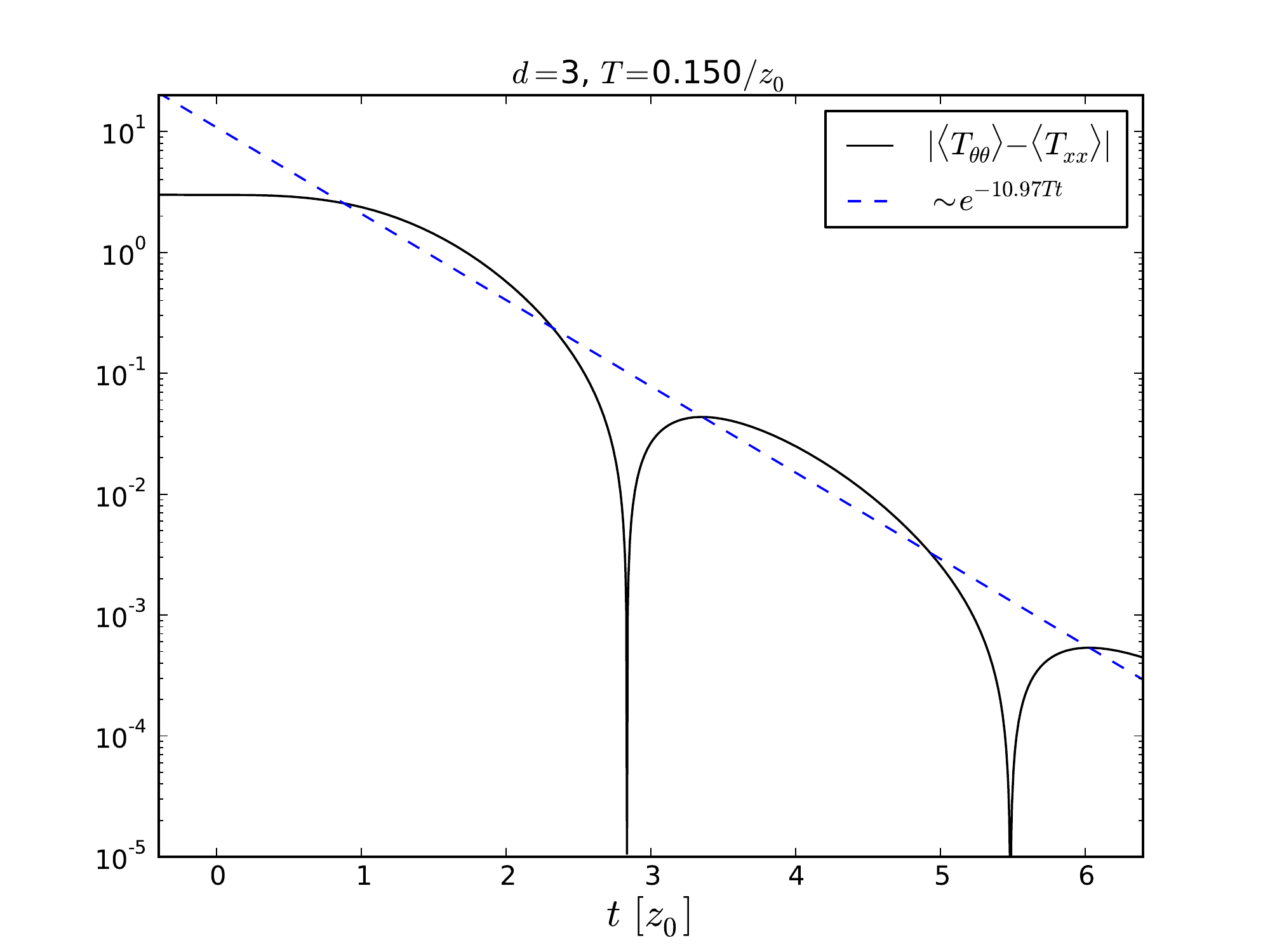}
\includegraphics[scale=0.33]{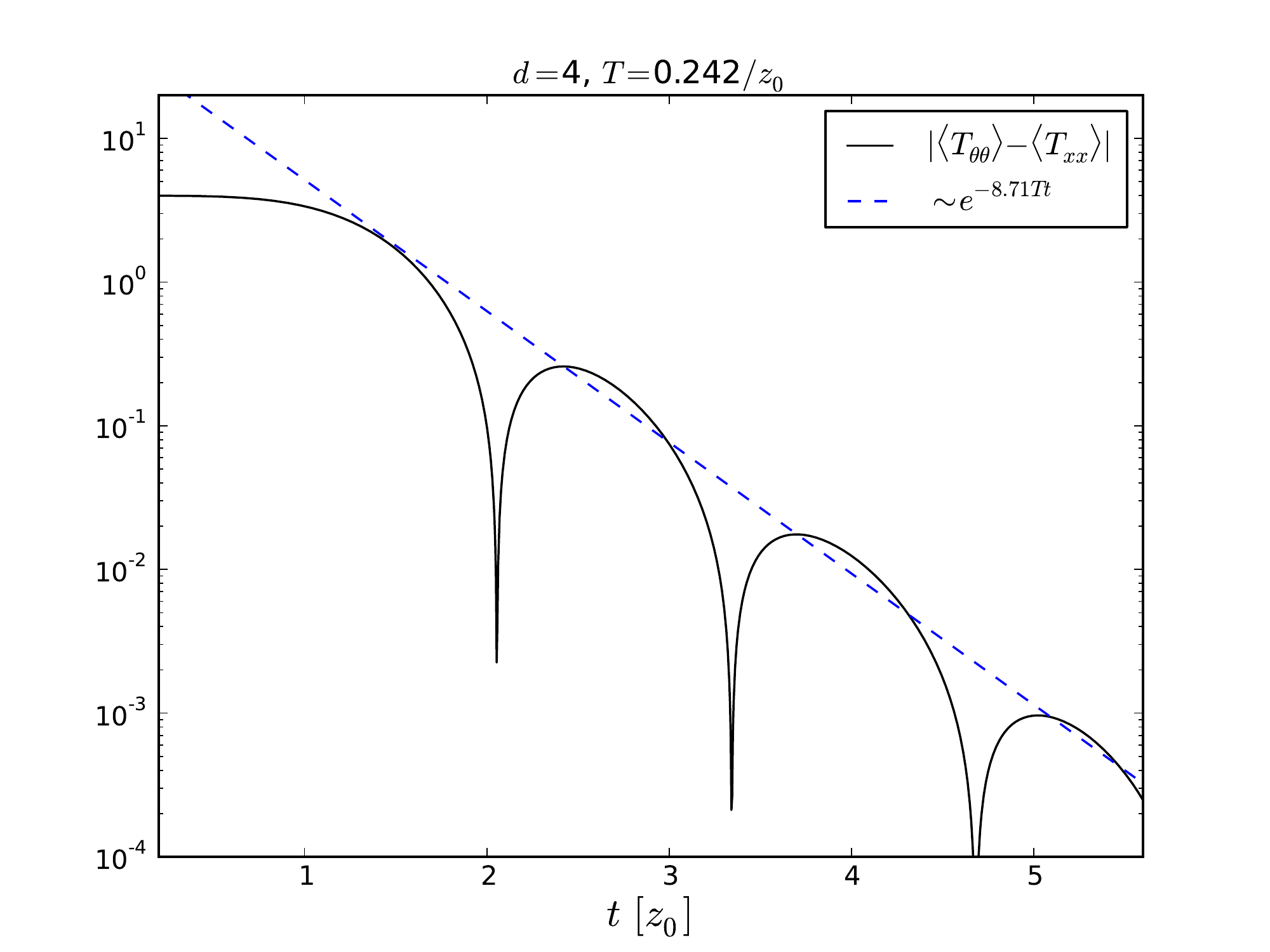}
\includegraphics[scale=0.33]{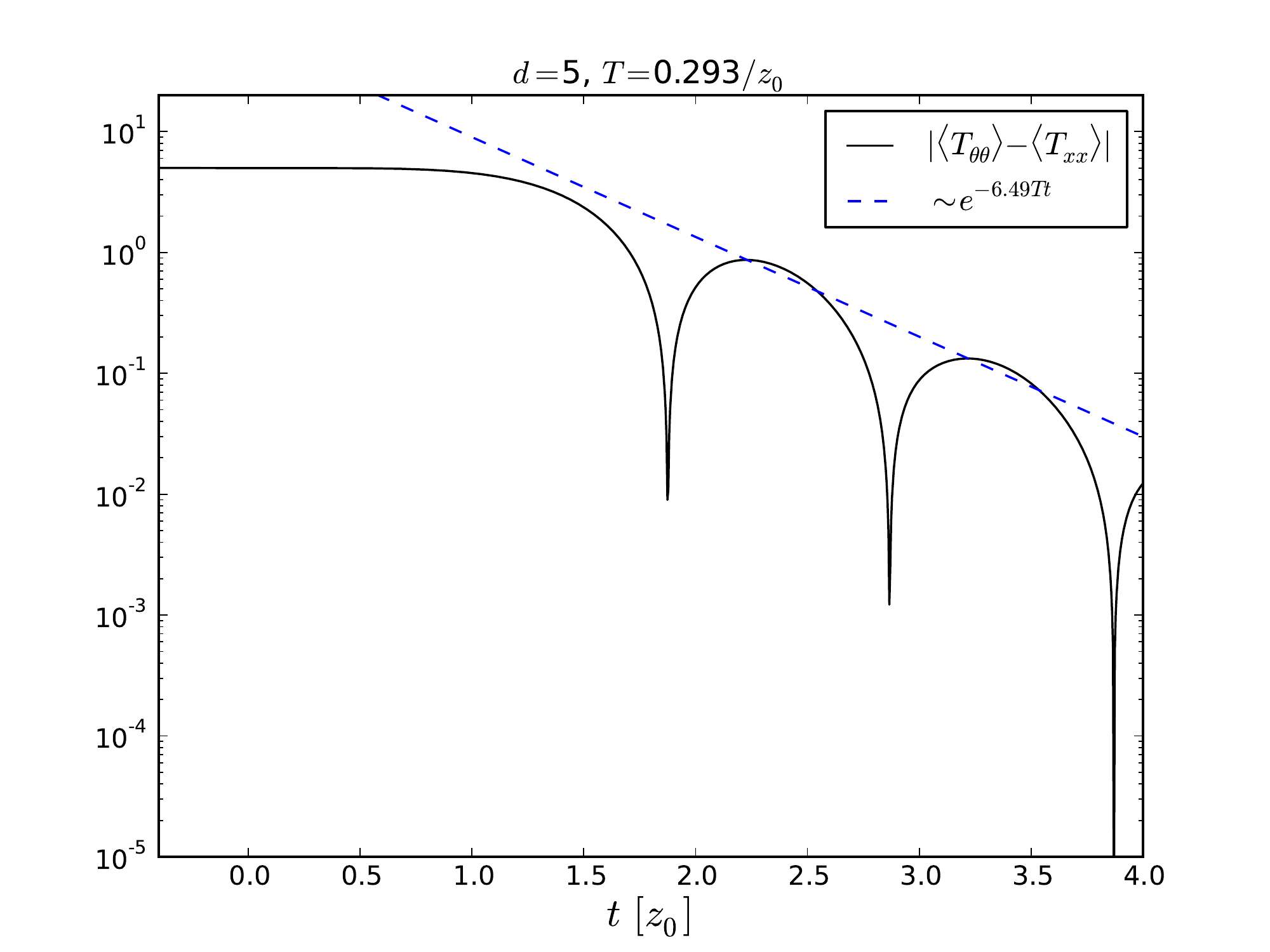}
\includegraphics[scale=0.33]{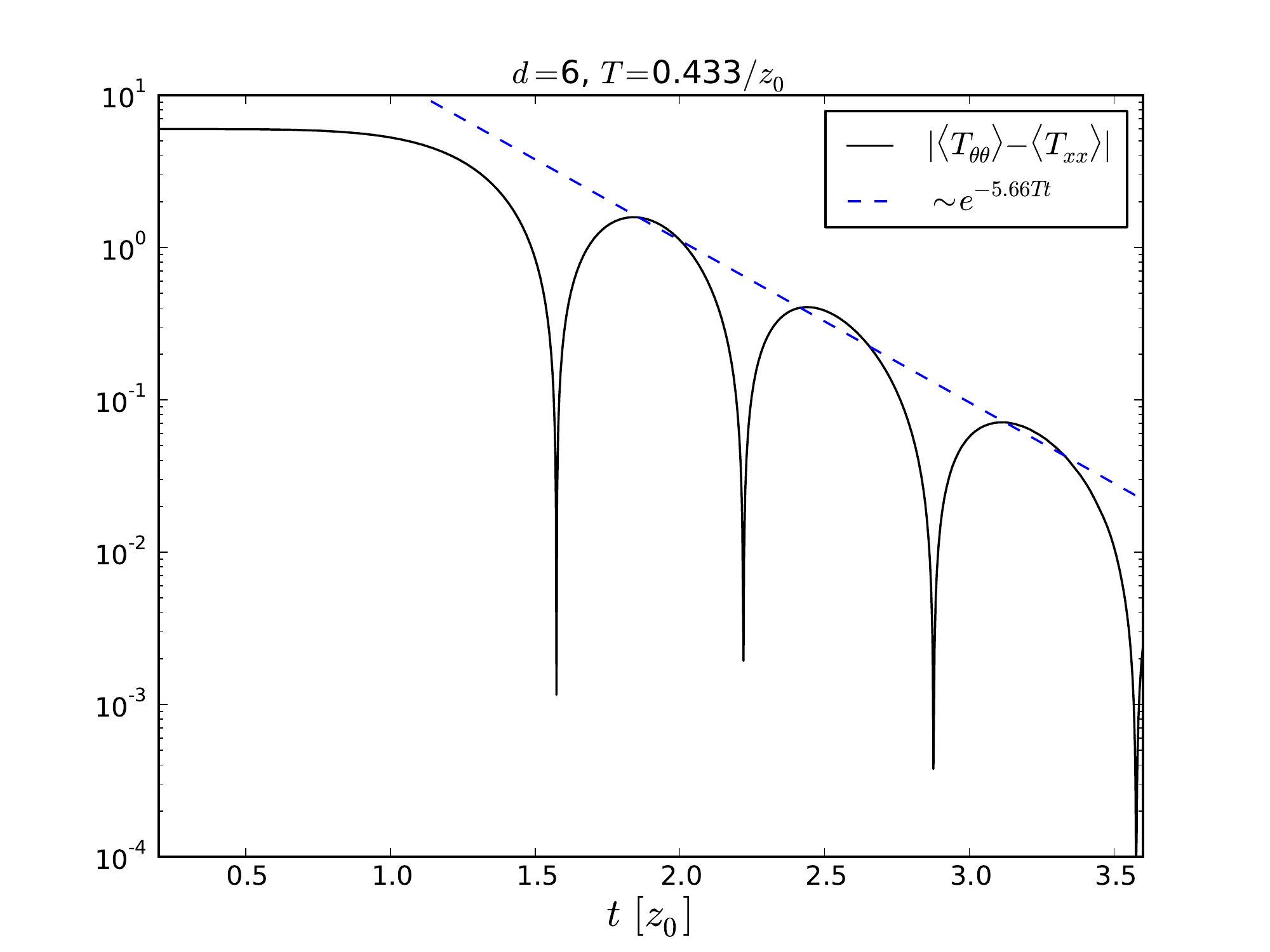}
\caption{Log scaled plots of the absolute deviations from zero of the pressure differences in the late time regime of some black hole collapse processes for various dimensions. Time is expressed in units of $z_0$, $\delta t=0.1z_0$, and the vacuum expectation values are given in units of $1/(16\pi G_N z_0^d)$. The decay constants $10.97T$, $8.71T$ and $5.66T$, where $T$ is the temperature, are in good agreement with the lowest quasinormal modes ($11.16T$, $8.63T$ and $5.47T$, respectively) quoted in \cite{Horowitz:1999jd}.}
\label{qnm} 
\end{figure}

For late times we can view the solution as being composed of a black brane background with small fluctuations. The late-time relaxation is thus expected to be governed by the lowest lying quasinormal modes for this black brane. A standard way to illustrate this behaviour is to plot the logarithm of the absolute value of the deviation of some observable from its final value. In Fig.~\ref{qnm}, a few examples of the deviation of the pressure difference $\langle T_{\theta\theta} \rangle-\langle T_{xx} \rangle$ from 0 are shown. We see that, as expected, the decay time is set by the lowest quasinormal mode, since the decay constants $10.97T$ (for $d=3$), $8.71T$ (for $d=4$) and $5.66T$ (for $d=6$) are in good agreement with the values for the lowest quasinormal mode frequencies of AdS Schwarzschild black branes obtained in \cite{Horowitz:1999jd}, namely $11.16T$, $8.63T$ and $5.47T$, respectively. 

\section{Scattering phase}\label{scatteringphase}
In the scattering phase, the scalar field wave packet that falls from the boundary, will bounce in the deep IR and return to the boundary. When it reaches the boundary there will typically be some excitation of the boundary observables. The wave packet then reflects from the boundary and the scattering repeats. There will be a similar quasiperiodic behavior in the metric, since due to the broken rotational symmetry between the $\vec{x}$ and $\theta$ coordinates the metric has dynamical degrees of freedom of its own. For all figures we have varied the grid spacing to make sure that the results are not numerical artifacts.

\begin{figure}[H]
\centering
\includegraphics[scale=0.6]{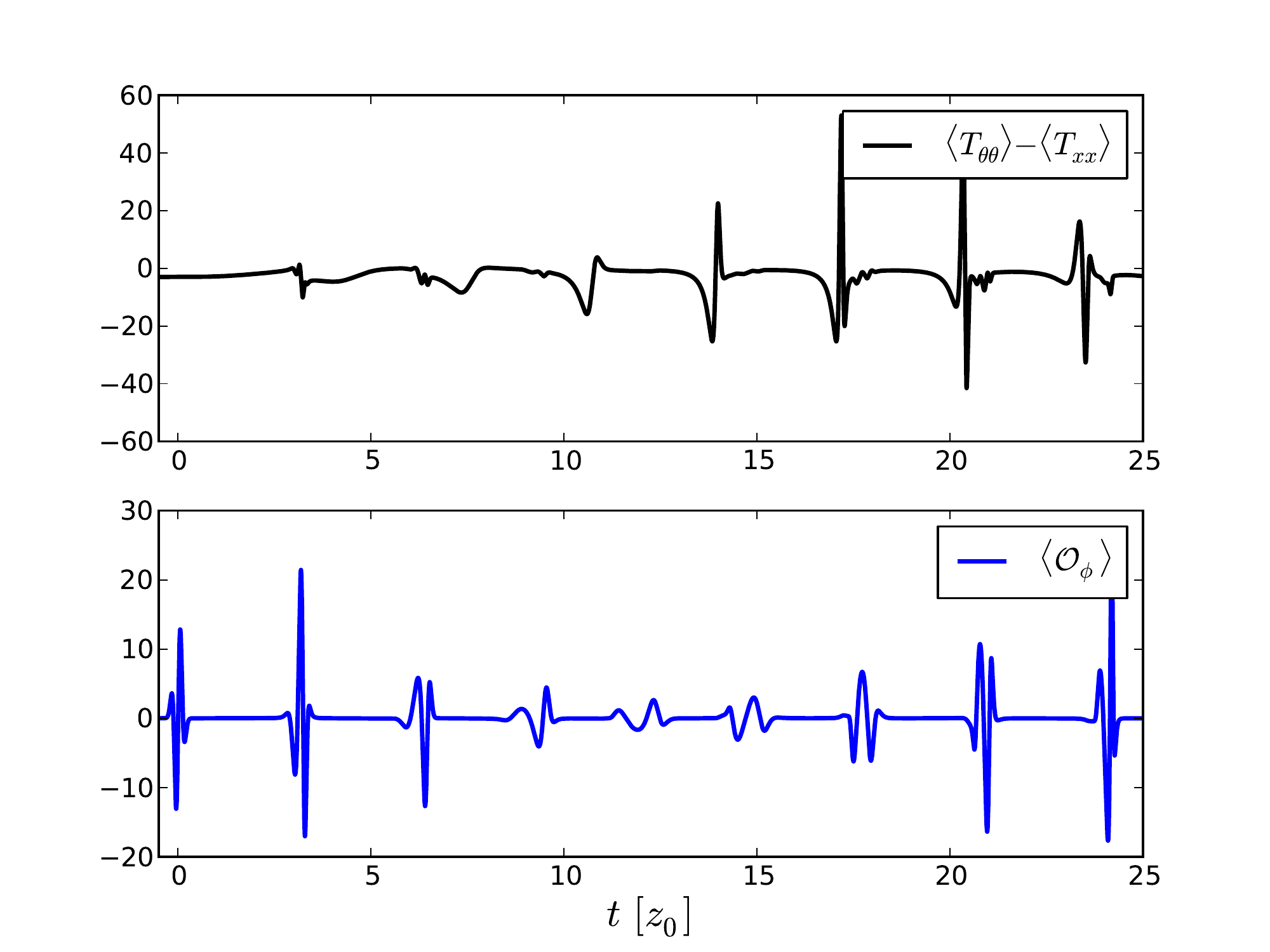}
\caption{The pressure components and vacuum expectation value of the scalar operator (in units of $1/(16\pi G_Nz_0^d)$)  for a scattering solution in $d=3$ with parameters $\epsilon=0.01$ and $\delta t=0.1z_0$.}
\label{scattering} 
\end{figure}

In Fig.~\ref{scattering}, we show a typical scattering solution. As we can see, every time the scalar field wave packet reaches the boundary, there is a bump in the expectation value, and this oscillation goes on forever as far as we know. We can also see that the dynamical degrees of freedom in the metric are excited, as expected, leading to non-trivial behavior in the boundary pressure components. 

\begin{figure}[H]
\centering
\includegraphics[scale=0.6]{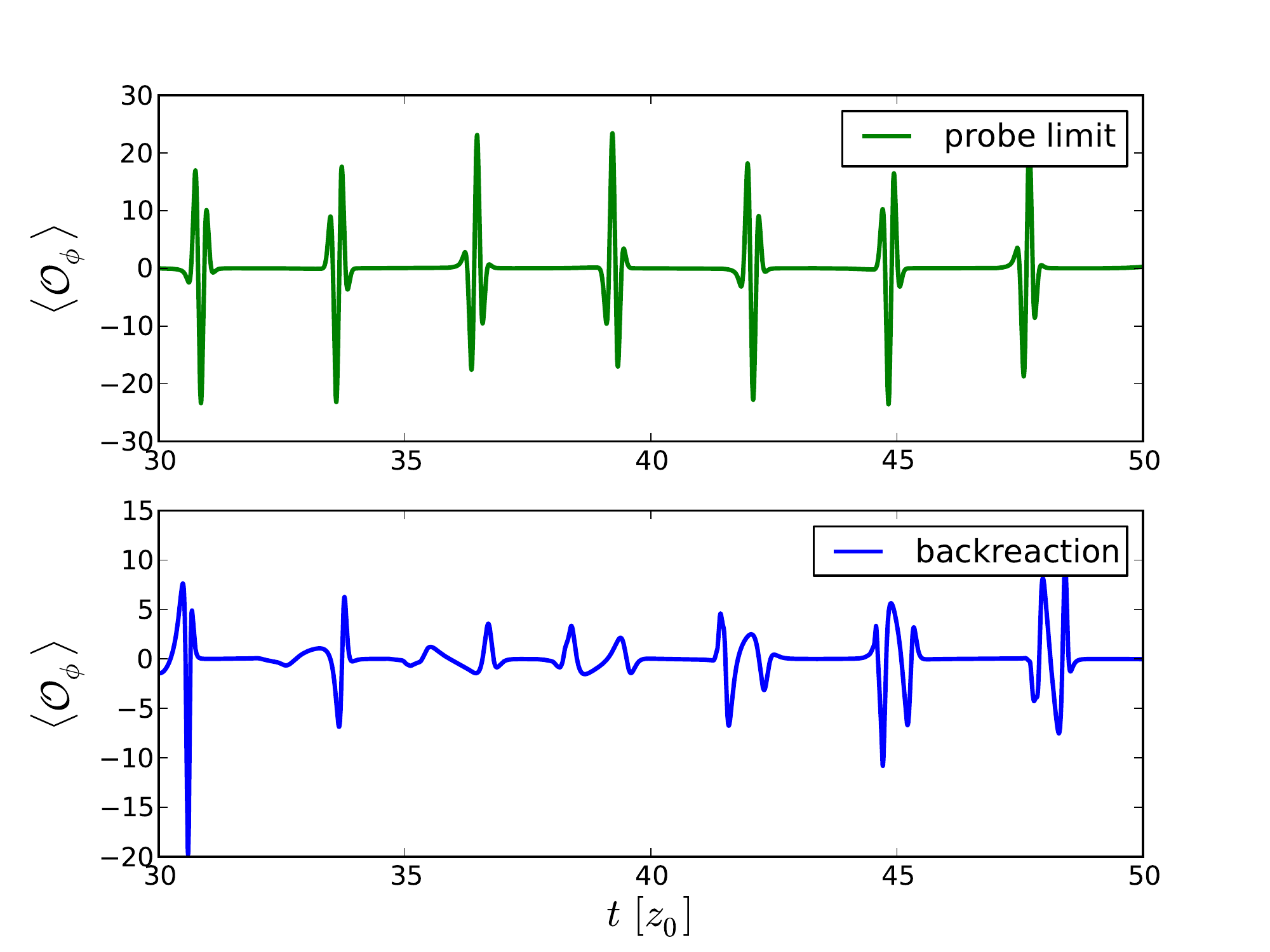}
\caption{The scalar expectation value compared with the probe limit result (in units of $1/(16\pi G_Nz_0^d)$). There is a clear distortion of the wave packet which is due to the non-trivial dynamics in the full Einstein equations and can not be seen in the probe limit, although the wavepacket remains fairly localized. This example is for $d=3$, with parameters $\epsilon=0.01$ and $\delta t=0.1z_0$, and time is in units of $z_0$. Note that this is already quite far into the non-linear regime since black hole formation occurs around $\epsilon\approx0.016$.}
\label{waveform} 
\end{figure}

One interesting feature is that the interpretation of the scattering solution as a localized wave packet persists for very long times, even for solutions where the non-linearities play a significant role. This is not obvious; one could have imagined that the wave packet would broaden and that at late times we would have seemingly random fluctuations, but instead we see that the wave packet remains approximately localized for long times. However, the shape of the wave packet can change with time due to the non-trivial dynamics of the full Einstein equations, as is reflected in Fig.~\ref{waveform} for a scattering solution close to the black hole threshold.

\begin{figure}[H]
\centering
\includegraphics[scale=0.6]{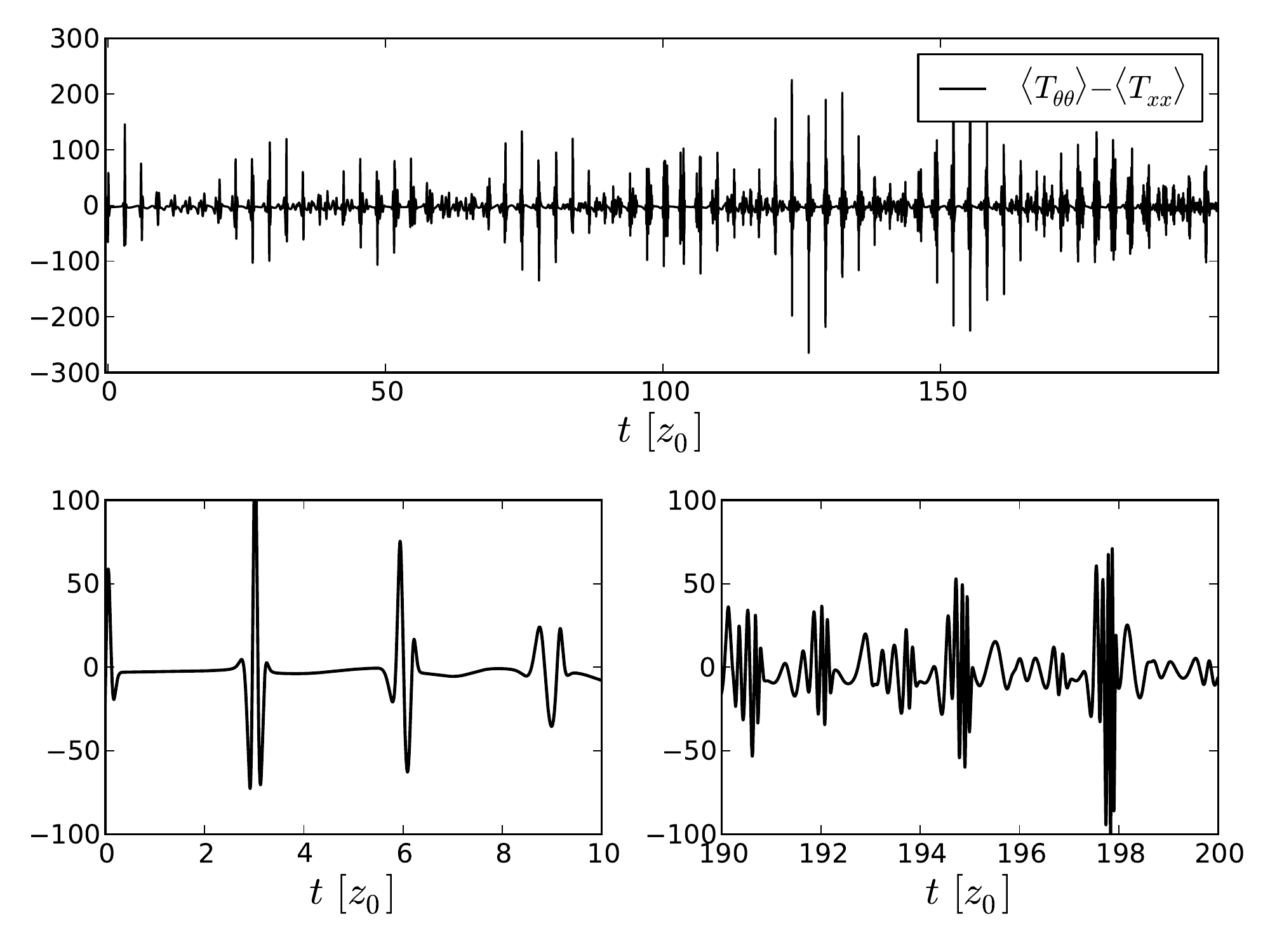}
\caption{The pressure difference after quenching the metric (in units of $1/(16\pi G_Nz_0^d)$), with $\epsilon=0.008$ and $\delta t=0.1z_0$ for $d=3$. Time is in units of $z_0$. We see signs of transfer to high frequency modes for late times.}
\label{gwscattering} 
\end{figure}

\begin{figure}[H]
\centering
\includegraphics[scale=0.6]{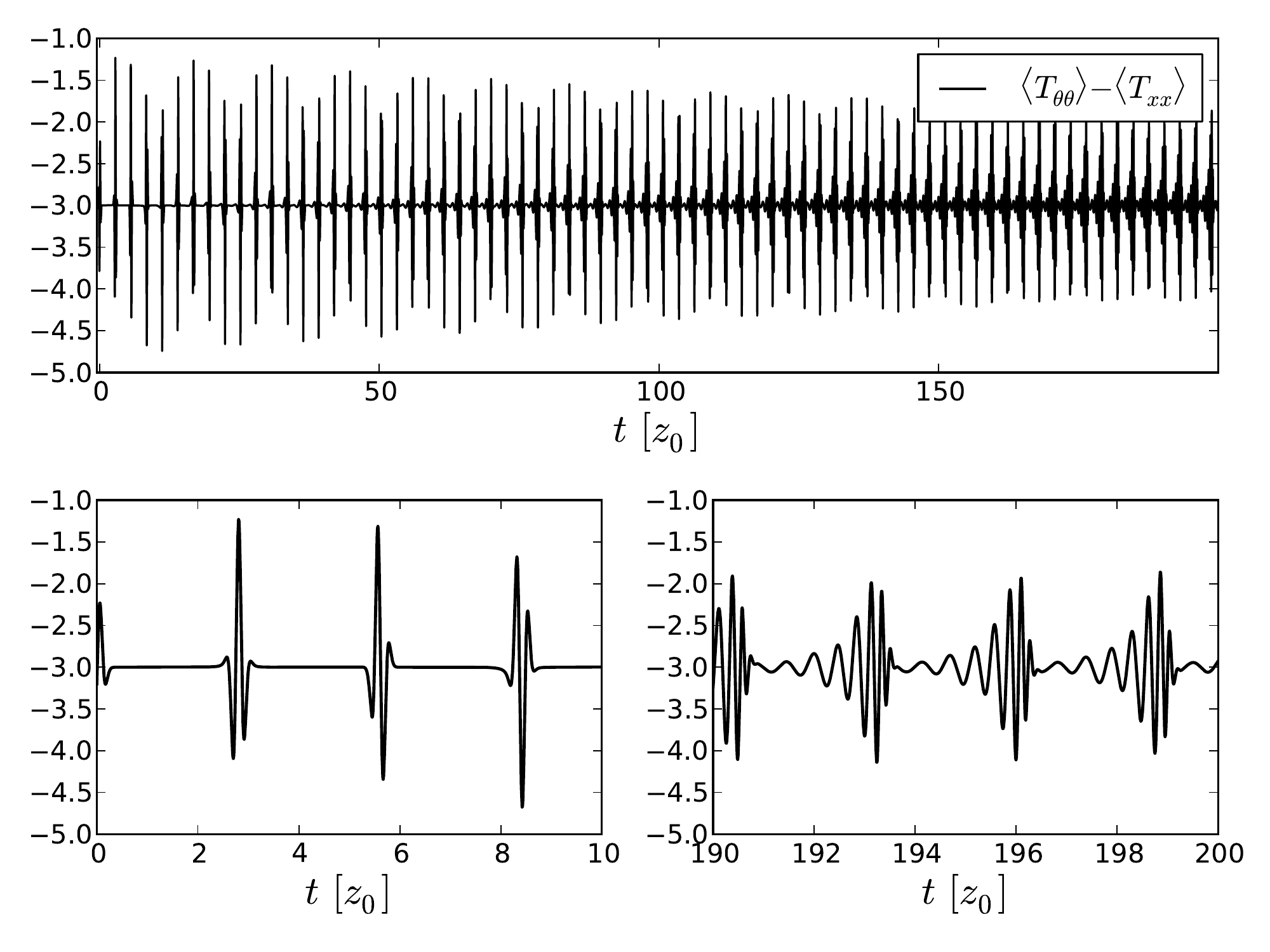}
\caption{The pressure difference after quenching the metric (in units of $1/(16\pi G_Nz_0^d)$), with $\epsilon=10^{-4}$ and $\delta t=0.1z_0$ for $d=3$. Time is in units of $z_0$. This figure is unchanged (except for an overall rescaling of the deviations from $-3$), when $\epsilon$ is decreased further. For these small amplitudes, we find no significant transfer to high frequency modes.}
\label{gwscattering001} 
\end{figure}

In Fig.~\ref{gwscattering} we show a typical long time scattering solution when the metric is quenched according to (\ref{bsource}). We notice that the pressure anisotropy develops increasingly sharp features after long times, which suggests transfer of energy to high frequency modes. At first sight, this might seem reminiscent of what happens to small-amplitude spherical scalar perturbations in  global AdS$_3$ \cite{Bizon:2013xha}, where turbulent transfer of energy to short wavelengths was interpreted as an instability of AdS$_3$. In Fig.~\ref{gwscattering001} we repeat our analysis for a smaller-amplitude source. While a Fourier analysis of the early and late time behavior in Fig.~\ref{gwscattering} confirmed the transfer of energy to higher frequencies, a similar analysis for Fig.~\ref{gwscattering001} showed no significant transfer to higher frequencies in the time range studied: in the latter case, the spectrum is dominated by normal mode frequencies, with roughly the same strength at early and late times. Decreasing the amplitude of the source further would simply rescale the vertical axis in Fig.~\ref{gwscattering001}, showing that for small amplitude the dynamics we see happens on timescales independent of the amplitude. The limited time range of our numerical simulation does not allow us to exclude transfer of energy on longer time scales (e.g., scaling as the inverse amplitude). However, based on the absence  of resonances in the normal mode spectrum in our setup, we expect no such energy transfer for small amplitudes. If so, the energy transfer observed in Fig.~\ref{gwscattering} is the result of strong nonlinearity and quite different from that of \cite{Bizon:2013xha}.

One important question is whether or not these scattering solutions will go on forever, or whether the system will approach some static solution. In section~\ref{static} we will show that, if the injected energy density is below the black brane threshold, the system must keep scattering forever.

\subsection{Static solutions and non-thermalization}\label{static}
 From equation \eqref{Ttt} we see that we are not able to form a black brane if $E<1/z_0^d$. However, one could imagine that there are other static solutions that the system can end up in. We will in this section show that if $E<1/z_0^d$ there are no static solutions that can be obtained through time evolution. To summarize the argument, the key information we get from the dynamical equations is the relation \eqref{feB2}. This condition is essentially the requirement that the spacetime should be regular at $z=z_0$ (such that a conical singularity can not be formed at this point during time evolution). We will then consider static solutions, by looking at the static equations of motion, and show that any possible static solution is incompatible with \eqref{feB2}. Actually, most of the solutions have a completely different asymptotic behavior at $z=z_0$ and are trivially excluded. The only solutions for which $fe^{-\frac{d-2}{2}b}$ goes to a constant, turns out to be the AdS soliton solutions. However, as we 
will see, all AdS solitons except our initial condition soliton will have $fe^{-\frac{d-2}{2}b}$ approaching a different constant than 1, violating \eqref{feB2}, so if the injected energy is non-zero, no static solutions can form. This reasoning is reminiscent of the argument for non-thermalization in the hard wall model, given in \cite{Craps:2014eba}, where we also had a relation similar to \eqref{feB2}.

To investigate this we will start by considering a different coordinate system, such that the metric takes the form
\begin{equation}
\rd s^2=\frac{1}{\hz^d}(-\hat{h}^2\rd t^2+\rd \hat{z}^2 \hat{f}^2+\rd \theta^2 e^{(d-2)\hat{b}}+\rd \vec{x}_{d-2}^2e^{-\hat{b}}). \label{newansatz}
\end{equation}
This can be obtained by the coordinate transformations and field redefinitions given by
\begin{equation}
\begin{array}{ccc}
b=\hb-\frac{\log G}{d-1}, &\hspace{10pt} z=\hz G^{\frac{1}{2(d-1)}},\\
\hf^2=\frac{f^2}{G(1-z \frac{G'}{2(d-1)G})^2}, &\hspace{10pt} \hh^2=h^2G^{-\frac{1}{d-1}},\\
\end{array}\label{transformations}
\end{equation}
 where $G(z)=1-z^d/z_0^d$ and $G'(z)=-dz^{d-1}/z_0^d$. The $\hz$ coordinate now ranges from $0$ to $\infty$. The (static) equations of motion for such an ansatz are
\begin{equation}
\frac{\hh'}{\hh}=\frac{\hf'}{\hf}-\frac{d}{\hz}(\hf^2-1),\label{hfeq1}
\end{equation}
\begin{equation}
\frac{\hh'}{\hh}=-\frac{\hf'}{\hf}-\frac{d-2}{4}\hz\hat{B}^2-\frac{1}{2(d-1)}\hz\hat{\Phi}^2,\label{hfeq2}
\end{equation}
\begin{equation}
(\frac{\hat{B}\hat{h}}{\hat{f}\hz^{d-1}})'=0,\label{Bseq}
\end{equation}
\begin{equation}
(\frac{\hat{\Phi}\hat{h}}{\hat{f}\hz^{d-1}})'=0,\label{Phiseq}
\end{equation}
where $\hB=\hat{b}'$, $\hat{\Phi}=\phi'$ and prime now denotes derivative with respect to $\hz$. We can integrate \eqref{Bseq} and \eqref{Phiseq} to obtain
\begin{equation}
\hat{\Phi}=C_\phi \frac{\hat{f}}{\hat{h}} \hz^{d-1},
\end{equation}
\begin{equation}
\hat{B}=C_b \frac{\hat{f}}{\hat{h}} \hz^{d-1}.
\end{equation}
where $C_b$ and $C_\phi$ are integration constants, tuning the UV behavior. From \eqref{hfeq1} we have $\hat{f}/\hat{h}=e^{\int_0^{\hz} \frac{d}{\hz'}(\hat{f}^2-1)}$, so that we obtain the following formulas for $\hat{B}$ and $\hat{\Phi}$
\begin{equation}
\hat{\Phi}=C_\phi e^{\int_0^{\hz} \frac{d}{\hz'}(\hat{f}^2-1)} \hz^{d-1},\label{Phisol}
\end{equation}
\begin{equation}
\hat{B}=C_b e^{\int_0^{\hz} \frac{d}{\hz'}(\hat{f}^2-1)} \hz^{d-1}.\label{Bsol}
\end{equation}
By eliminating $\hat{h}'/\hat{h}$ from \eqref{hfeq1} and \eqref{hfeq2} and substituting the expressions in \eqref{Bsol} and \eqref{Phisol} for $\hat{B}$ and $\hat{\Phi}$ we obtain that $\hat{f}$ must satisfy
\begin{equation}
2\frac{\hf'}{\hf}-\frac{d}{\hz}(\hf^2-1)=-\frac{d-2}{4}C_b^2\hz^{2d-1}e^{\int_0^{\hz}\frac{2d}{\hz'}(\hf^2-1)\rd \hat{z}'}-\frac{1}{2(d-1)}C_\phi^2\hz^{2d-1}e^{\int_0^{\hz}\frac{2d}{\hz'}(\hf^2-1)\rd \hz'}.\label{staticsolutions}
\end{equation}
With the boundary expansion of $f$ being $f=1+\frac{E}{2(d-1)}z^d+\ldots$ (see \eqref{fUV}), we obtain the boundary expansion of $\hf$ as $\hf=1+\frac{(E-1/z_0^d)}{2(d-1)}\hz^d+\ldots$. We expect that black branes will form when the total energy density is positive, which from \eqref{Ttt} corresponds to $E-1/z_0^d>0$. Here we will now consider the case $E<1/z_0^d$ (negative energy density) and show that any possible static solutions with negative energy density cannot be obtained dynamically. We emphasize that some solutions of \eqref{staticsolutions} might have singular behaviours and should be excluded as relevant solutions by other arguments, but we will not care about such issues here, and just directly show that any static solutions, which must satisfy \eqref{staticsolutions}, cannot be formed dynamically with the AdS soliton as initial condition. Recall the relation \eqref{feB2}, which says that we must have $fe^{-\frac{d-2}{2}b}=1$ in the IR when $\hat{z}\rightarrow\infty$ or equivalently $z\
rightarrow z_
0$. The idea is now to show that all solutions obtained by solving \eqref{staticsolutions} are inconsistent with this requirement. We will use the notation $\approx$ to mean that two quantities are equal asymptotically, while $\sim$ means that they are equal asymptotically up to an overall constant.

To show this we will have to compute the IR asymptotic behaviour of the solutions \eqref{staticsolutions}. We first note that the derivative $\hat{f}'$ in \eqref{staticsolutions} is negative if $0<\hf<1$. Since $\hf=1+(E-1/z_0^d)\hz^d/2(d-1)+\ldots<1$ close to the boundary, we obtain that $\hf<1$ for all $\hz$.

It is also easy to see that $\hf$ cannot become negative (because \eqref{staticsolutions} implies that if $\hat{f}=0$ then around that point $\hat{f}\sim \hat{z}^{-\alpha}$ for an $\alpha>0$ so $\hat{f}$ can only go to zero asymptotically). Also, $\hat{f}$ can not asymptote to any other constant than zero. This can be seen by assuming that $\hat{f}\rightarrow c>0$, and then \eqref{staticsolutions} implies that $\hat{f}\sim e^{-\alpha\hz^{\beta}}$ for some $\alpha,\beta>0$ which is inconsistent with $\hat{f}\rightarrow c$ (unless if $C_\phi=C_b=0$, in which case $\hat{f}\sim \hat{z}^{-\alpha}$ for $\alpha>0$, which is also inconsistent, or if $\hat{f}\equiv1$). Since $\hat{f}$ is strictly decreasing, it thus follows that we must have $\hf\rightarrow0$ when $\hz\rightarrow\infty$. When $\hz\rightarrow\infty$ we thus have that
\begin{equation}
e^{\int_0^{\hz}\frac{d}{\hz'}(\hf^2-1)\rd \hat{z}'}\approx C' \hz^{-2d}
\end{equation}
for some constant $C'$. For simplicity of notation we can thus redefine $C_b C'=C_{b, IR}$ and $C_\phi C'=C_{\phi, IR}$. The asymptotic behaviour of $\hat{b}$ is then
\begin{equation}
\hat{b}\approx C_{b,IR}\log \hz. \label{bIR}
\end{equation}
Equation \eqref{staticsolutions} now becomes in the IR
\begin{equation}
 2\frac{\hf'}{\hf}\approx-\left(\frac{d-2}{4}C_{b,IR}^2+\frac{1}{2(d-2)}C_{\phi,IR}^2+d\right)\hz^{-1},
\end{equation}
from which we can obtain the asymptotic behaviour
\begin{equation}
\hat{f}\sim \hat{z}^{-\frac{d-2}{8}C_{b,IR}^2-\frac{1}{4(d-2)}C_{\phi,IR}^2-\frac{d}{2}}.\label{hatfIR}
\end{equation}
We must now compute the asymptotic relations between ($f$, $b$) and ($\hat{f}$, $\hat{b}$), by using the expressions in \eqref{transformations}. We have that $\hz\approx z_0G^{-1/(2(d-1))}$, which directly implies the asymptotic relations
\begin{equation}
b\approx \hat{b}+2\log\frac{\hz}{z_0}\ \ 
\end{equation}
and
\begin{equation}
f\approx \frac{d}{2(d-1)}\hat{f}\left(\frac{\hz}{z_0}\right)^{d-1},\label{IRfehatf}
\end{equation}
which imply
\begin{equation}
fe^{-\frac{d-2}{2}b}\approx \frac{d}{2(d-1)}\frac{\hz}{z_0}\hat{f}e^{-\frac{d-2}{2}\hat{b}} .\label{IRfeb}
\end{equation}
From the above relations and \eqref{bIR} and \eqref{hatfIR} we now obtain the asymptotic behaviour
\begin{equation}
b\approx(C_{b,IR}+2)\log \hz,\label{BIR}
\end{equation}
\begin{equation}
f^2\sim \hz^{\left(2(d-1)-\frac{d-2}{4}C_{b,IR}^2-\frac{1}{2(d-2)}C_{\phi,IR}^2-d\right)},
\end{equation}
so that we finally obtain
\begin{equation}
fe^{-\frac{d-2}{2}b}\sim\hz^{-\frac{d-2}{2}(\frac{C_{b,IR}}{2}+1)^2-\frac{1}{4(d-2)}C_{\phi,IR}^2}.\label{IRfe}
\end{equation}
We thus see that $fe^{-\frac{d-2}{2}b}$ will generically vanish in the IR, trivially violating the condition that $fe^{-\frac{d-2}{2}b}\rightarrow1$. The only way to have $fe^{-\frac{d-2}{2}b}$ approach a constant in the IR, is when the power in \eqref{IRfe} vanishes. This only happens when $C_{\phi,IR}=0$ and $C_{b,IR}=-2$, which in particular implies that the scalar identically vanishes. Only for these particular IR parameters will $fe^{-\frac{d-2}{2}b}$ go to a constant. We will now show, however, that it will go to a constant different from 1.

To specify a solution in the bulk, it would be customary to specify the UV behavior, meaning that we specify $E$ and $C_{b}$ and then integrate to the IR, which should give a unique solution. Specializing to $C_{b,IR}=-2$ should leave us a one parameter family of static solutions. Below we will construct this one parameter family of solutions, which turns out to be all the AdS solitons.

An AdS soliton solution with a general confinement scale $z_1$, can be given by the metric \eqref{zansatz} with $b=0$ and $f=h=1$ with $z_0$ replaced by $z_1$. After transforming to the metric \eqref{newansatz}, by using the transformations in \eqref{transformations}, we can obtain the asymptotic behavior for $\hf$ and $\hat{b}$ as $\hat{b}\approx-2 \log \frac{\hz}{z_1}$ and $\hf\approx\frac{2(d-1)}{d} \left(z_1/\hz\right)^{d-1}$. 
We can now easily obtain from \eqref{IRfeb} that $fe^{-\frac{d-2}{2}b}\rightarrow z_1/z_0$. Thus, the only possible solution that can be obtained is $z_1=z_0$ which corresponds to our initial condition, and which corresponds in the UV to $E=0$. 

To conclude, when the total energy density lies between that of the AdS soliton and zero (the threshold for black brane formation), no static solutions exists.

\subsection{Long time amplitude modulation} \label{sec:long}

\begin{figure}[H]
\centering
\includegraphics[scale=0.33]{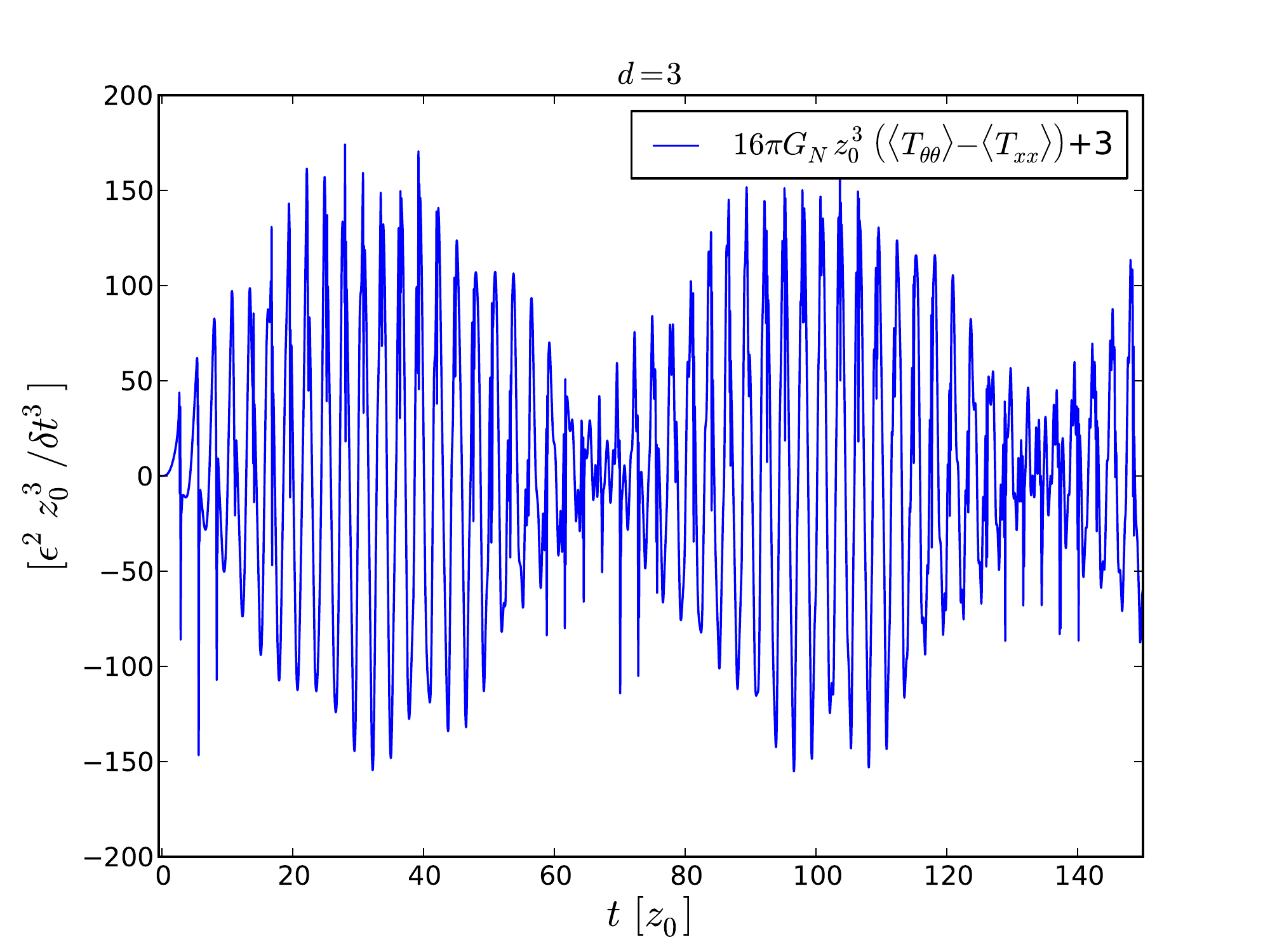}
\includegraphics[scale=0.33]{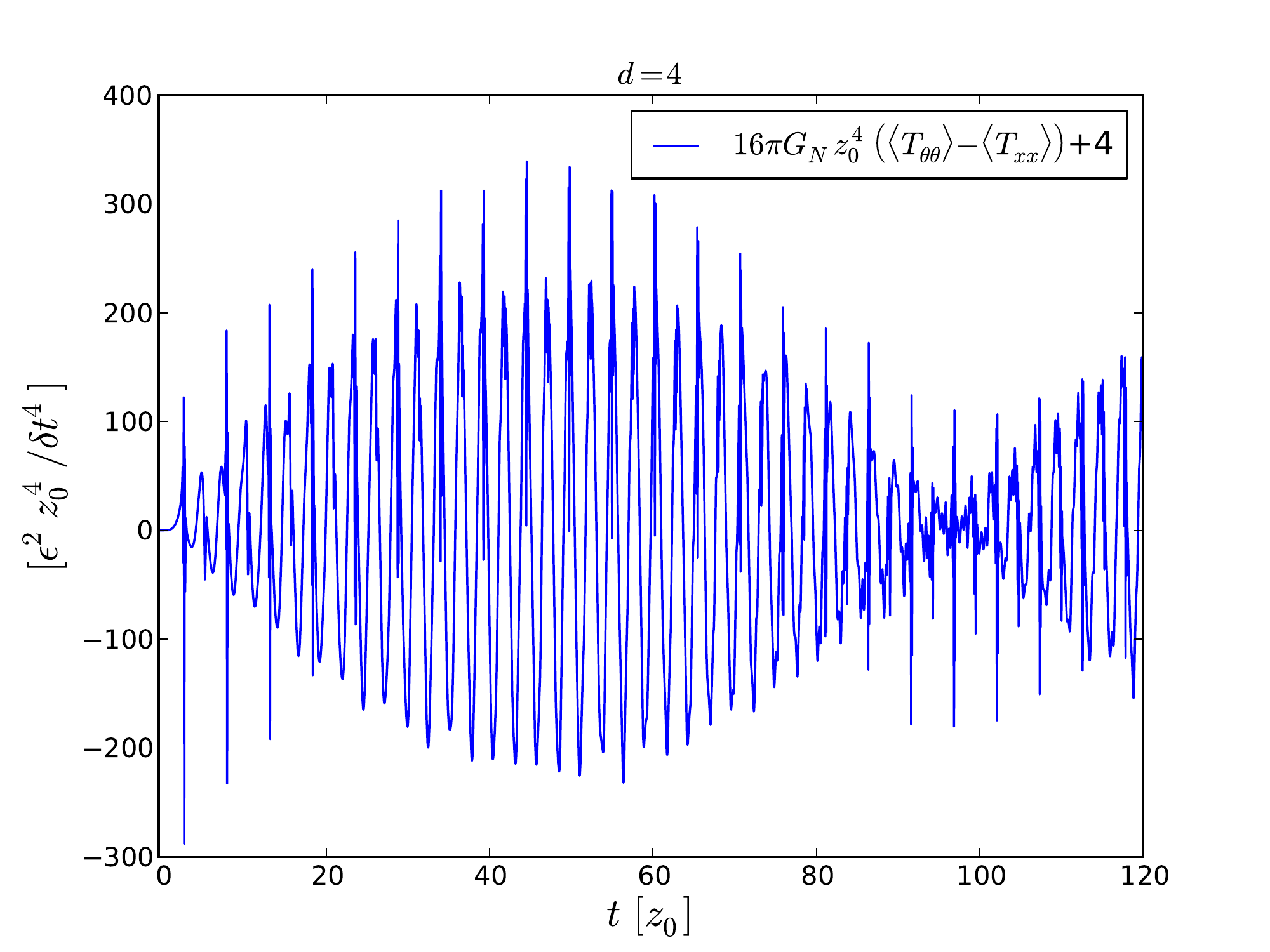}
\includegraphics[scale=0.33]{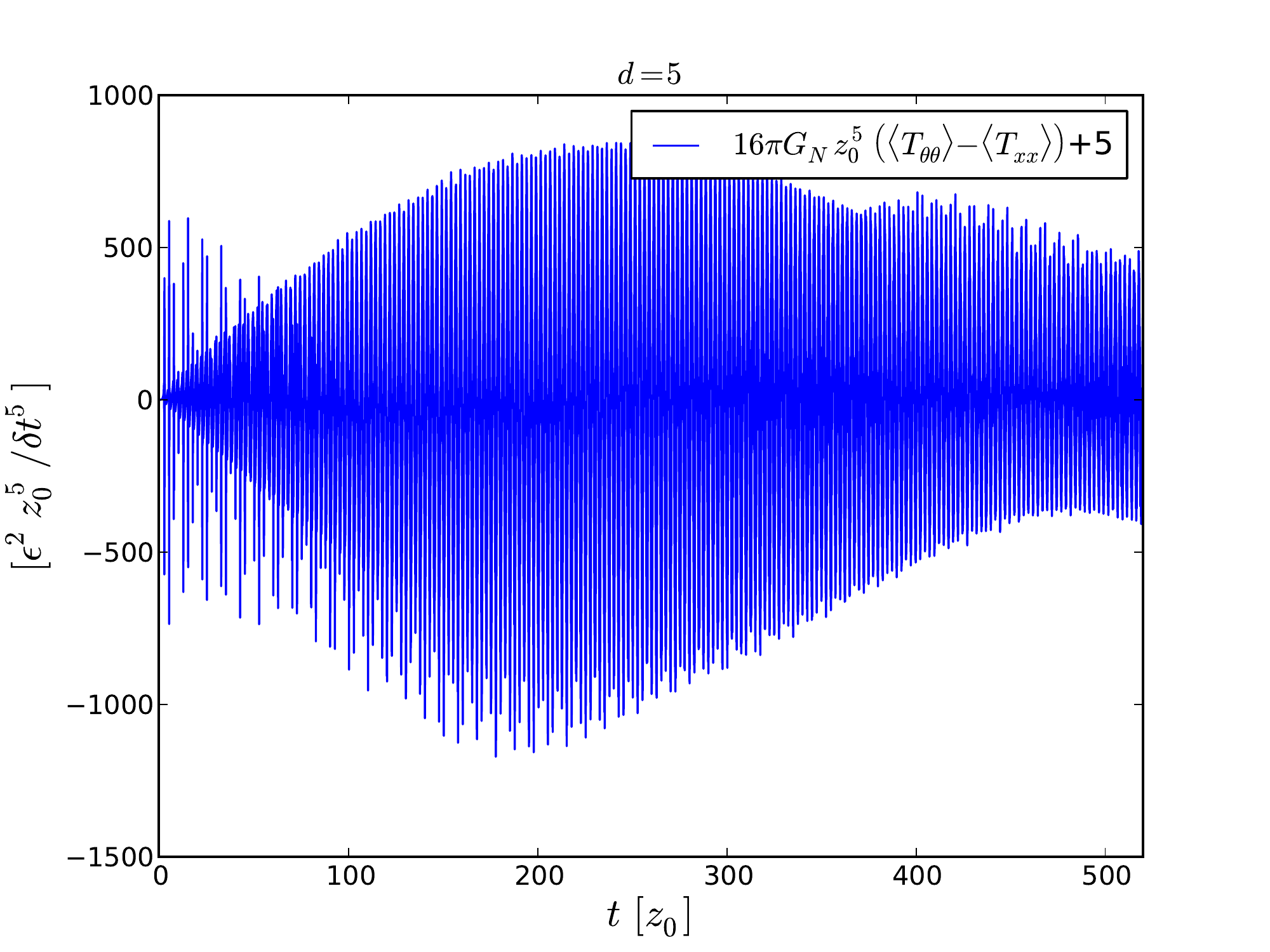}
\includegraphics[scale=0.33]{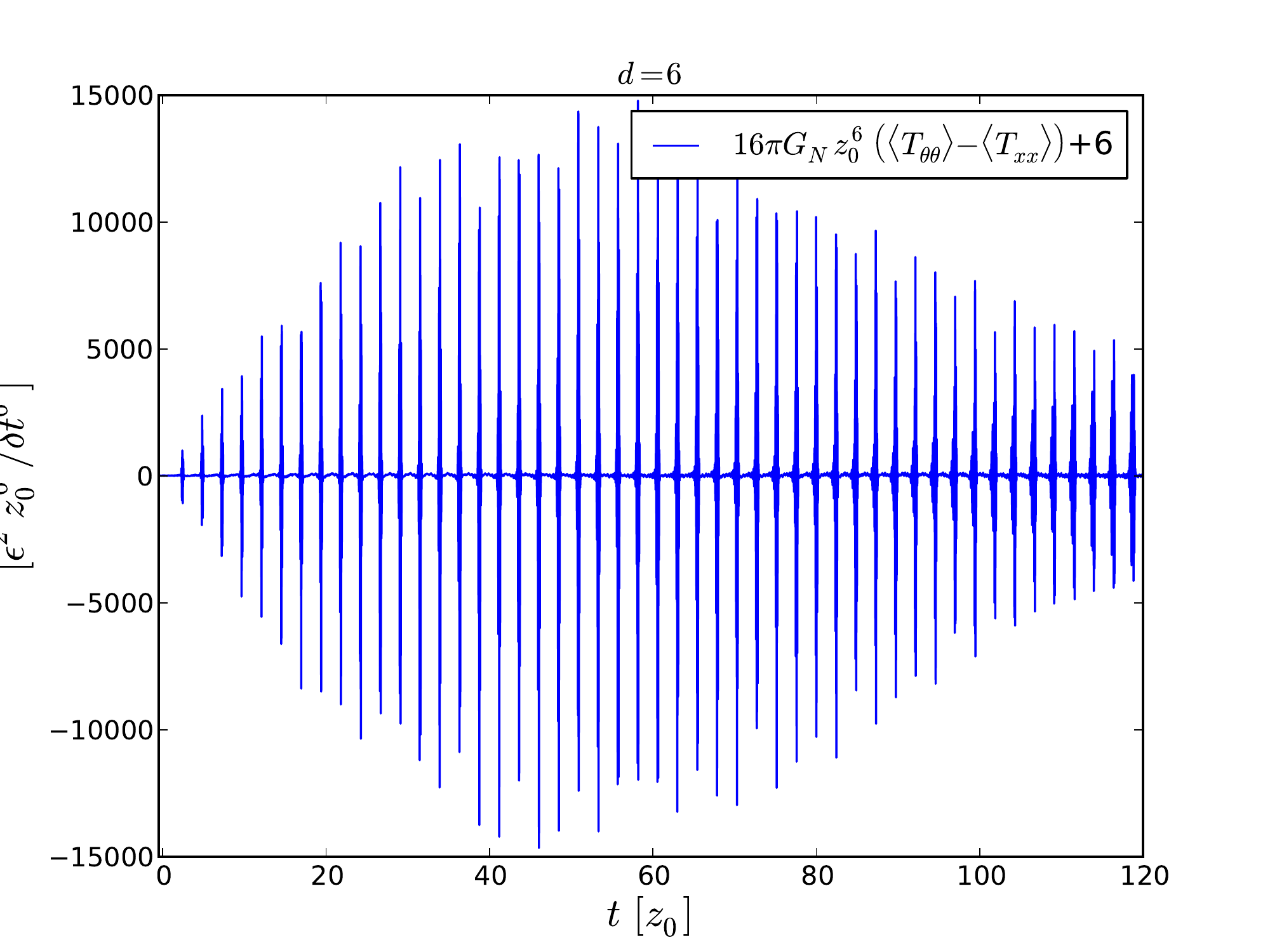}
\caption{The pressure anisotropy after a weak scalar perturbation with small $\epsilon$ and $\delta t=0.1z_0$. Time is measured in units of $z_0$, and the vertical axis has been rescaled by $\epsilon^2/\delta t^d$, which is the expected dependence of the total energy of the system for small $\epsilon,\delta t$. For $d=3,4,5$ we see that the amplitude undergoes an amplitude modulation on a much longer time scale which is in excellent agreement with the result in table \ref{modtable}. For $d=6$, the modulation due to the first normal mode is hidden by the peaks from the bouncing wave packet part; it is clearly visible, however, in the normal mode decomposition in Fig.~\ref{normalmode_decomposition}.}
\label{amplitudemod} 
\end{figure}

For small-amplitude scattering solutions (small $\epsilon$), we observe an amplitude modulation in the pressure anisotropy on a long time scale, see Fig.~\ref{amplitudemod}. (The relevant timescale is actually hard to see for $d=6$, for reasons we will explain below.) The time scale can be seen to be independent of $\epsilon$ and $\delta t$ as long as both parameters are sufficiently small. This is different from the $1/\epsilon^2$ modulation time scale observed in \cite{Craps:2014eba} for the $d=3$ hard wall model with Neumann boundary conditions. As we will now explain, in the present case the modulation is due a near-resonance between a metric mode and the bouncing scalar shell. (In the $d=3$ hard wall model with Neumann boundary conditions, no dynamical metric modes were excited, and the modulation was due to a resonant spectrum of scalar field normal mode frequencies.)

In the small $\epsilon$ regime, the scalar field is $\mathcal{O}(\epsilon)$. Thus the metric is of order $\mathcal{O}(\epsilon^2)$ and the next order corrections to the scalar are $\mathcal{O}(\epsilon^3)$. Working to order $\mathcal{O}(\epsilon^2)$, we can therefore consider $\phi$ as a probe scalar acting as an external source on the metric. Since the scalar $\phi$ bounces back and forth between the IR and the boundary, the source for the metric backreaction can be characterized by a frequency\footnote{Not to be confused with the metric component $f$ from previous sections.} 
$f_\phi$. In the limit of small $\delta t$ (the thin shell limit), this frequency will be the same as for a lightlike particle (following lightlike geodesics) that bounces between the boundary and the interior. In particular, for small $\delta t$ the bounce frequency becomes independent of $\delta t$. 

\begin{figure}[H]
\centering
\includegraphics[scale=0.33]{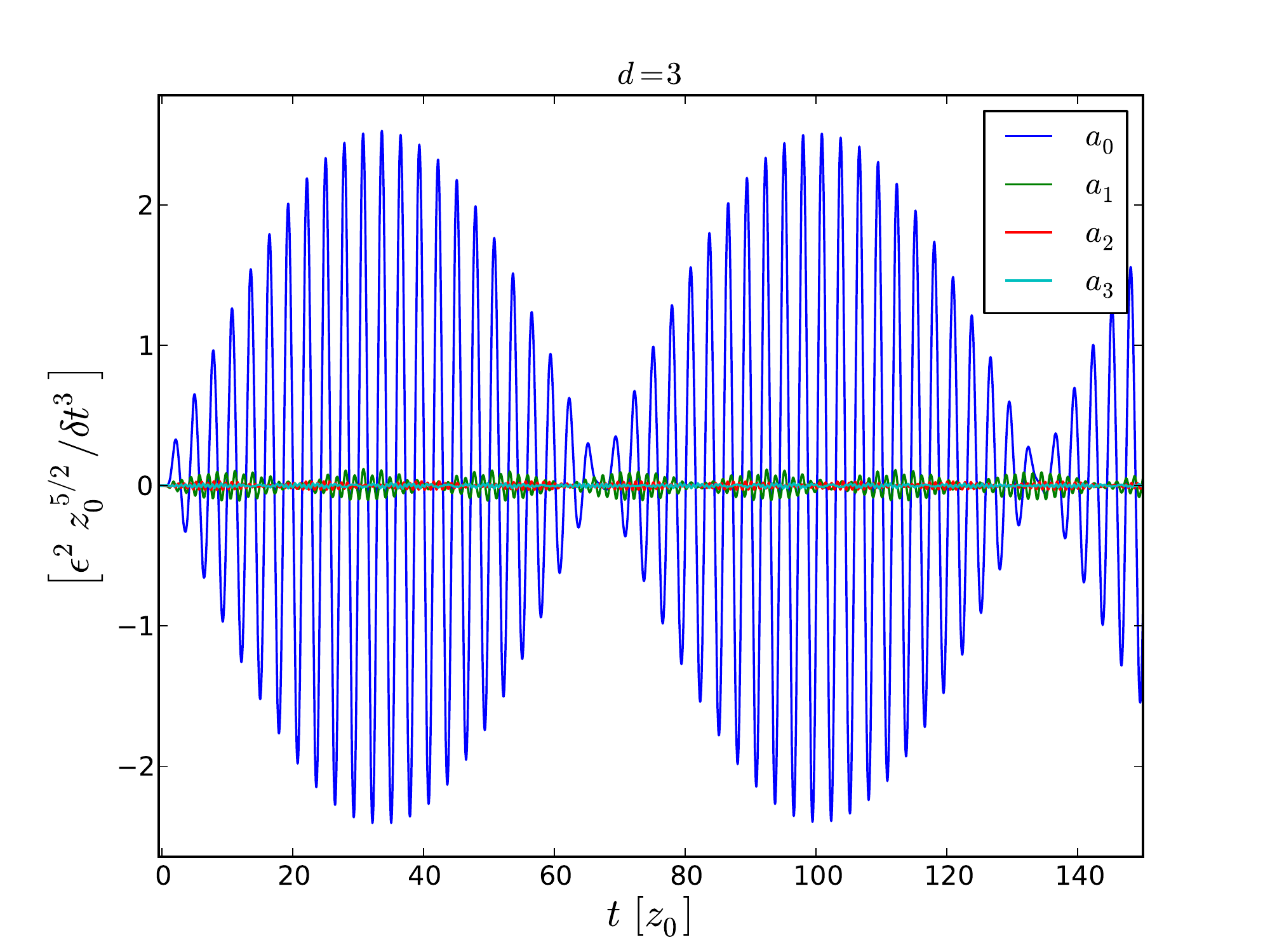}
\includegraphics[scale=0.33]{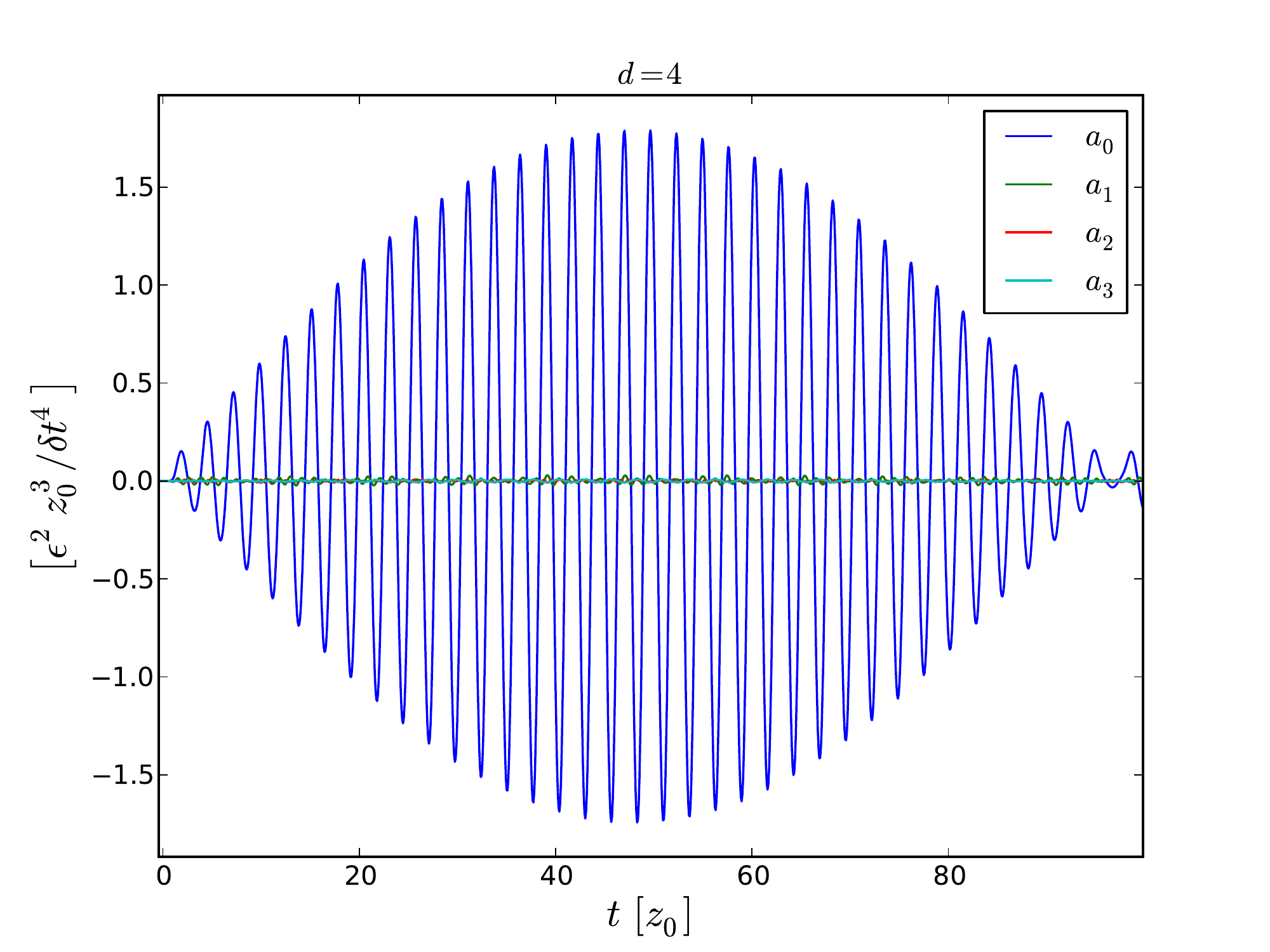}
\includegraphics[scale=0.33]{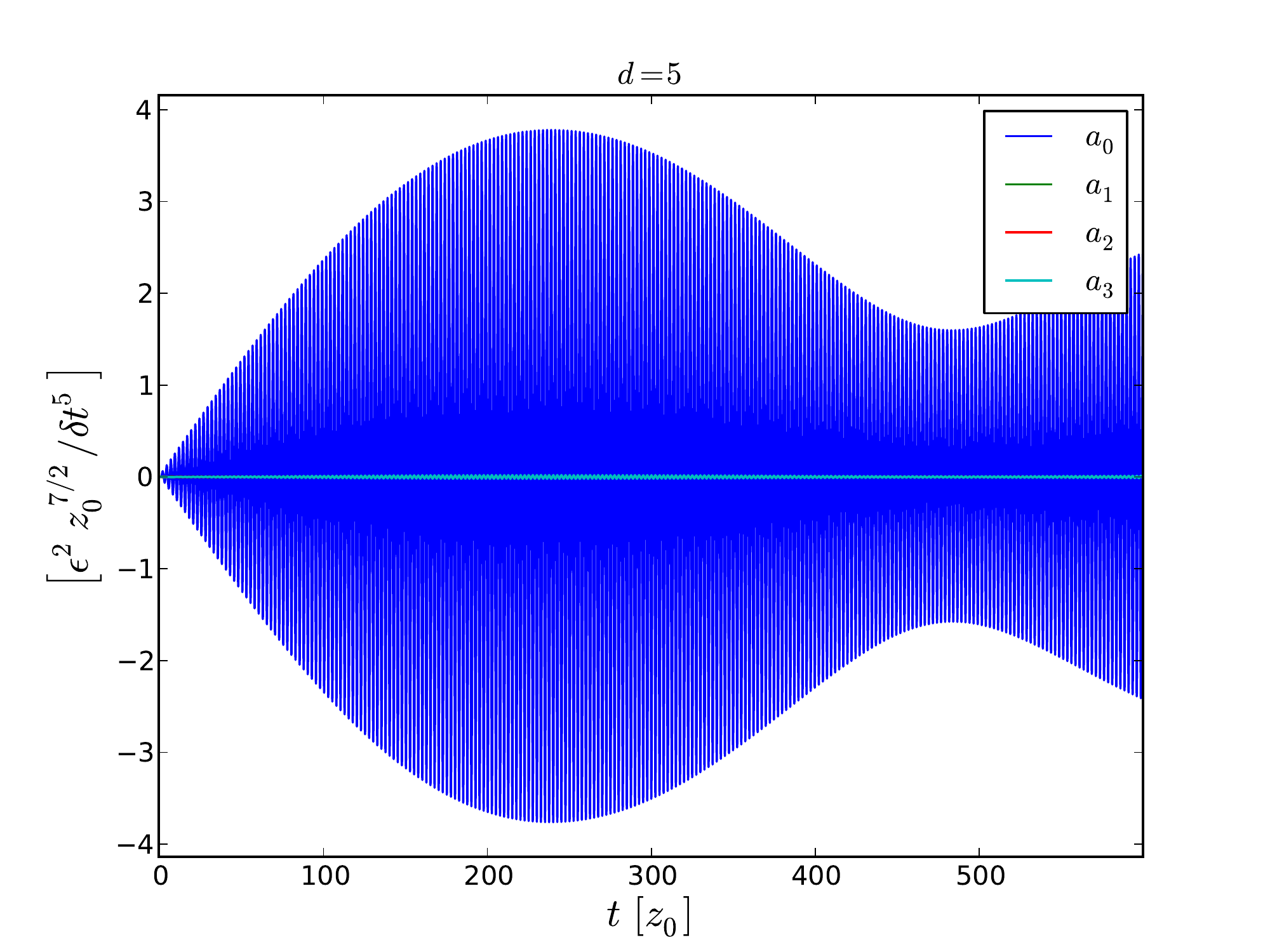}
\includegraphics[scale=0.33]{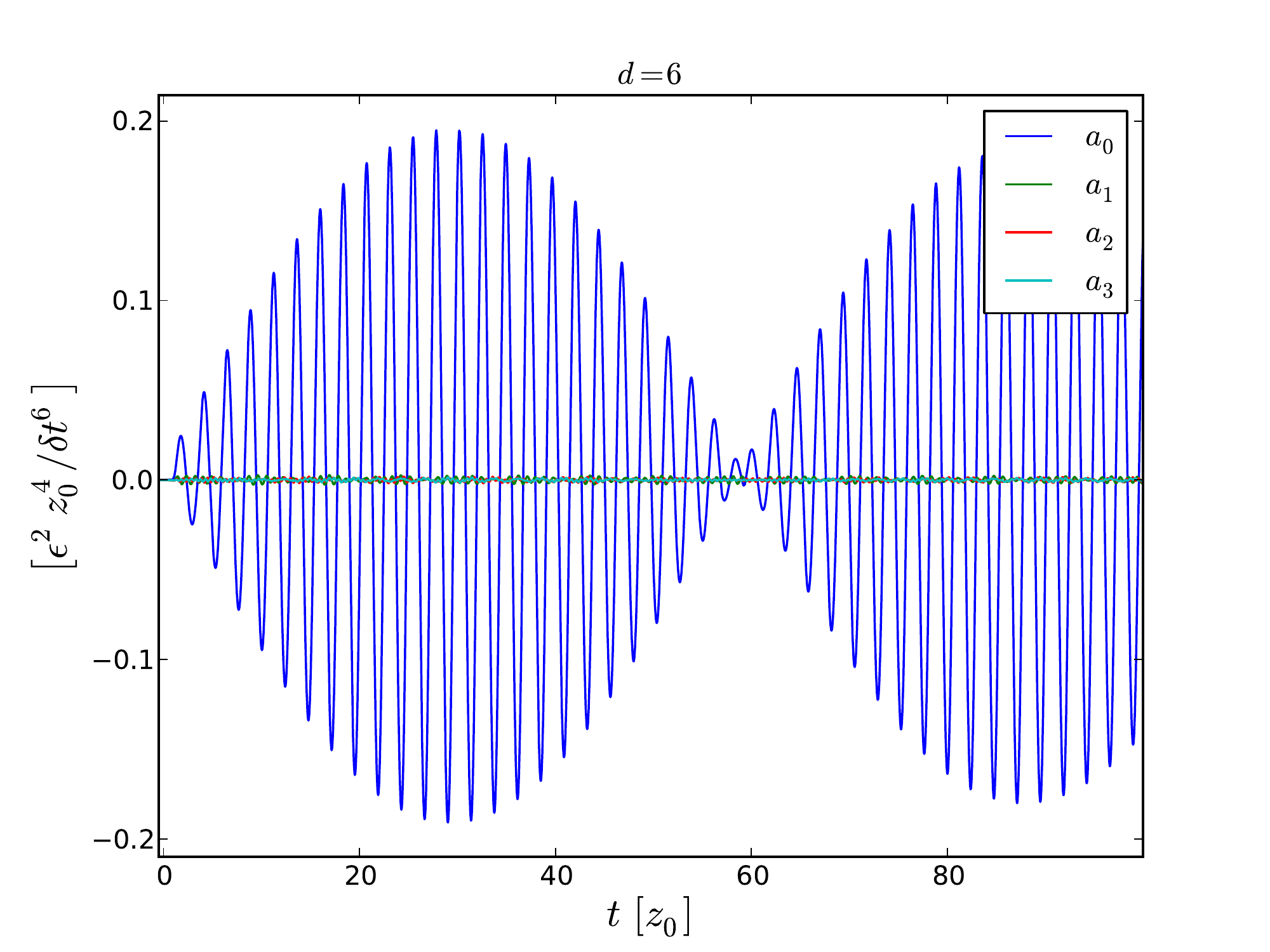}
\caption{The metric function $b$ decomposed in normal modes, after a weak scalar perturbation with small $\epsilon$ and $\delta t=0.1z_0$. Time is measured in units of $z_0$. We see that, as expected, the lowest mode is more excited than higher modes, and undergoes an amplitude modulation which agrees with the result in Table~\ref{modtable}.}
\label{normalmode_decomposition} 
\end{figure}

However, the metric also has some intrinsic frequencies $f_i$ (the normal mode frequencies, see Appendix \ref{normalmodes}). Every time the scalar crosses the space-time it kicks the metric. It is useful to decompose the metric fluctuation into its normal modes. If $f_\phi=f_j$ for some $j$, we would expect a resonance, such that the amplitude of the $j$'th normal mode will increase linearly with time. But if $f_\phi\approx f_j$, such that we are close to resonance, it would be natural to expect another time scale showing up, namely $T=1/|f_\phi-f_j|$. The results are summarized in table \ref{modtable}, and can be compared with the numerical results in Fig.~\ref{amplitudemod} and Fig.~\ref{normalmode_decomposition}. The latter figure shows the decomposition of $b$ in its normal modes, where the decomposition is of the form $b(z,t)=\sum_{n\geq0} a_n(t)Q_n(z)$. The functions $Q_n$ (corresponding to the $n$th normal mode with frequency $\omega_n$) satisfy the equation \eqref{Qeqbis} with $\omega=\omega_n$, 
which constitutes a Sturm-Liouville problem (which makes this decomposition possible), and they are normalized with respect to the inner product $\int_0^{z_0} Q_n(z)Q_m(z)z^{-(d-1)}dz=\delta_{mn}$. Note that replacing the frequency of  the scalar wave packet by that of a light-like thin shell still gives decent result, but using the true frequency is required to get accurate results, especially for $d=5$ (where the system is very close to resonance). To summarize, the modulation can be traced back to a near resonance between the lowest normal mode frequency of metric perturbations and the frequency of a bouncing scalar shell. As expected from this picture, this type of modulation does not show up when we quench the metric instead of the scalar field, as can be seen in Fig.~\ref{gwscattering001}.

One can see from the numerics, however, that the metric perturbation consists not only of a few normal modes: it has a slowly moving normal mode part and a rapidly moving wave packet part. The wave packet part is in general smaller than the normal mode part. However, close to the boundary, the wave packet part can still give large contributions to the boundary observables. The intuitive explanation is as follows. Close to the boundary a wave packet looks typically like $~\psi((z-t)/\delta t)$, where $\psi$ is some localized profile and $\delta t$ is the width. Thus when extracting the $z^d$ coefficient when computing the boundary observables, this will be proportional to $\partial_z^d\psi((z-t)/\delta t)\sim 1/\delta t^d$, while the derivatives of the normal modes are of $\mathcal{O}(1)$. We thus see that for larger dimensions, the wave packet part is expected to become more important. These are exactly the sharp peaks one can see in  Fig.~\ref{amplitudemod}, and indeed they become larger for larger 
dimansions. For $d=6$ 
they completely dominate and this is the reason why we cannot see the modulation due to the first normal mode in the vacuum expectation value in $d=6$. However, it can still be seen in the normal mode decomposition in Fig.~\ref{normalmode_decomposition}, since here the contribution from the wave packet part is still small.

\begin{table}
\begin{center}
\begin{tabular}{ |c|c|c|c|c| } 
 \hline
 $d$ & 3 & 4 & 5 & 6\\ 
 \hline
 $z_0f_0$ & 0.34195 &0.37177  &0.40151&0.43004\\ 
 $z_0f$ & 0.35682(19) & 0.38190(36) &0.39944(16)&0.41263(43)\\ 
 $z_0f'$ & 0.3564 & 0.3807 &0.3986&0.4117\\ 
 $T/z_0$ & 67.2(8) & 98.7(35) &484(37)&57.4(14)\\ 
 $T'/z_0$ & 69.2 & 111.9 &344&54.5\\ 
 \hline
\end{tabular}
\end{center}
\caption{\label{modtable} The lowest normal mode frequency $f_0$, the oscillation frequency of the scalar wave packet $f$ for $\delta t=0.1z_0$, the oscillation frequency of a lightlike thin shell $f'$ and the expected modulation times $T=1/|f_0-f|$ and $T'=1/|f_0-f'|$ using $f$ respectively $f'$. Note that the frequency of a thin shell is extremely close to the frequency of the bouncing scalar field. However, note also that in $AdS_6$ ($d=5$) we are extremely close to resonance, and to get an accurate modulation time we must use the true frequency of the scalar wave packet (compare with Fig.~\ref{amplitudemod} and Fig.~\ref{normalmode_decomposition}). The estimated error comes from reading off the oscillation frequencies of the wave packet from the numerical simulations, while the errors of $f_0$ and $f'$ are negligible.}
\end{table}

\subsubsection{Harmonic oscillator toy model}\label{oscillator_example}
To develop a better understanding of the modulations we have just described, we now study a sourced harmonic oscillator which is conceptually similar to our gravitational setup (in the small-amplitude scattering phase) and which experiences a similar amplitude modulation phenomenon.

\begin{figure}[H]
\centering
\includegraphics[scale=0.7]{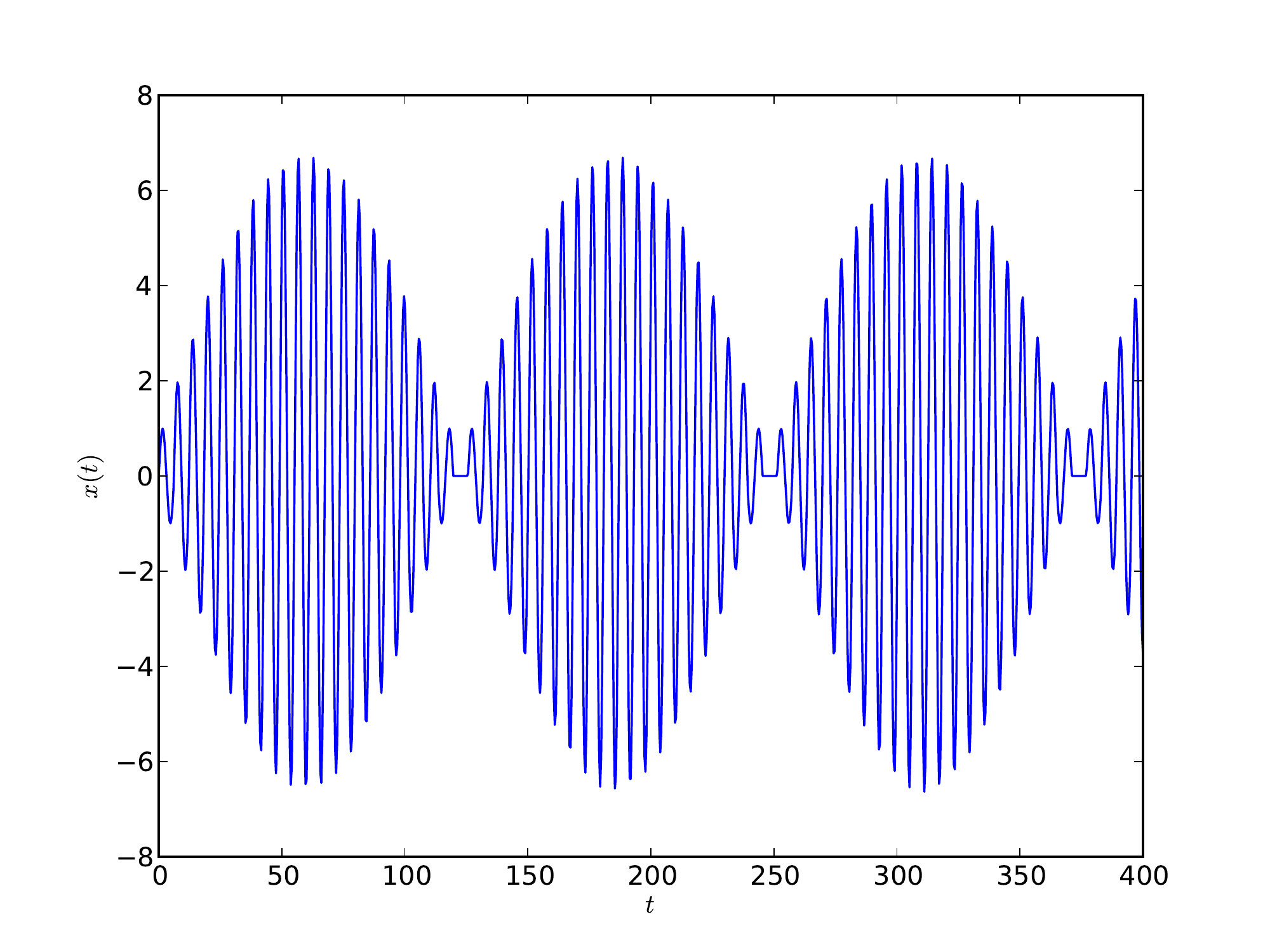}
\caption{Evolution of the sourced harmonic oscillator given by equation \eqref{oscillator}, close to resonance with parameters $\omega=1$ and $T=2\pi/1.05$. We see that the result is a rapidly oscillating solution with a long time modulation with period $2\pi/0.05\approx125.7$.}
\label{oscillator_modulation}
\end{figure}

Consider a harmonic oscillator with angular frequency $\omega$, sourced by a sequence of local kicks (modelled by delta functions) with period $T$. (We denote the frequency of the kicks by $f=1/T$.) The equation of motion is
\begin{equation}
\ddot{x}+\omega^2x=\sum_{n\geq0} \delta(t-nT),\label{oscillator}
\end{equation}
subject to the initial condition that $x(t)$ should vanish for $t<0$. To compare with our gravitational setup, $x$ is the analogue of $B$ (the metric backreaction), the delta functions are analogous to the stress energy tensor for the scalar $\phi$ which sources the metric, the frequency $f=1/T$ is analogous to the oscillation frequency $f_\phi$ of $\phi$, and $\omega$ is analogous to the normal mode frequencies of the metric perturbations. We can solve \eqref{oscillator} by performing a Laplace transform. For the Laplace transformed field $X$ we have
\begin{equation}
s^2X(s)+\omega^2X(s)=\sum_{n\geq0} e^{-nTs}\Rightarrow X(s)=\frac{1}{2i\omega}\sum_{n\geq0}\left(\frac{e^{-nTs}}{s-i\omega}-\frac{e^{-nTs}}{s+i\omega}\right).
\end{equation}
It is now easy to do the inverse Laplace transform, to obtain
\begin{equation}
x(t)=\sum_{n\geq0}\frac{\sin(\omega(t-nT))}{\omega}\theta(t-nT),
\end{equation}
where $\theta(t)$ is the Heaviside step function. By letting $N=\lfloor t/T \rfloor$ (the largest integer less than or equal to $t/T$), we can write this as
\begin{equation}
\omega x(t)=\sum_{n=0}^N\sin(\omega(t-nT))=\mathrm{Im }\enspace e^{i\omega t}\sum_{n=0}^Ne^{-i\omega nT}=\mathrm{Im }\enspace e^{i\omega t}\frac{1-e^{-i\omega (N+1)T}}{1-e^{-i\omega T}},\label{geosum}
\end{equation}
under the assumption that $T\omega\not\equiv 0 \pmod{2\pi}$. Extracting the imaginary part and using some trigonometric identities, we obtain
\begin{equation}
x(t)=\frac{2\sin\left[\frac{\omega T}{2}\right]\sin\left[\omega(t-\frac{NT}{2})\right]\sin\left[\omega\frac{N+1}{2}T\right]}{\omega(1-\cos\left[\omega T\right])}.\label{3sine}
\end{equation}
If the system is almost at resonance, $T\omega\approx2\pi$, the middle factor in \eqref{3sine} will give rise to fast oscillations, while the last factor gives rise to slow amplitude modulations. To see this, we write $f-\omega/2\pi=\epsilon\ll f$ (so $\epsilon$ is the difference between the source frequency and the oscillator frequency). The third factor in \eqref{3sine} now becomes
\begin{align}
\sin\left[\omega\frac{N+1}{2}T\right]&=\sin\left[\pi(1-\epsilon T)(N+1)\right]=\pm \sin \left[\pi \epsilon T (N+1)\right]\nonumber\\
&\approx\pm \sin \left[\pi \epsilon (t+T)\right],
\end{align}
where in the last step we approximated $N=\lfloor t/T \rfloor\approx t/T$, and we see that we indeed obtain an overall amplitude modulation with period $1/\epsilon$. An example with $\omega=1$ and $\epsilon=0.05/2\pi$ is shown in Fig.~\ref{oscillator_modulation}. Further, we note that it is the small denominator in (\ref{3sine}) that causes a near-resonant normal mode to dominate the other normal modes.

If $T\omega =2\pi k$, for some non-zero integer $k$, the summation of the geometric series in \eqref{geosum} yields instead
\begin{equation}
x(t)=\frac{N+1}{\omega}\sin\omega t\approx \frac{t+T}{2\pi k}\sin\omega t,
\end{equation}
which is a sine function with a (step-wise) increasing amplitude, as expected when we are at resonance. This can also be obtained as a limit $T\omega\rightarrow 2\pi k$ in \eqref{3sine}.

\section{Conclusions and outlook}\label{conclusions}

A common feature of the hard wall model and the AdS soliton is the energy gap between the ground state and the lightest black hole. This feature underlies the dynamical phase structure uncovered in  \cite{Craps:2013iaa, Craps:2014eba} (for the hard wall model) and in the present work (for the AdS soliton). 

In the black hole phase, we have found that the late-time decay is governed by the lowest quasinormal mode, as could have been expected. It was nice, however, to find numerical solutions in this model that collapse after one or more bounces, which is reminiscent of  spherical collapse in global AdS \cite{Bizon:2011gg} -- it would be interesting to study the similarities and differences in more detail. A good starting point could be the observation that the AdS$_3$ soliton is identical to global AdS$_3$.

In the scattering phase, we have established that, if the injected energy density is below the black hole threshold, relaxation to a static solution that is translationally invariant in the spatial directions along the boundary is impossible. An interesting question for future work is whether there are instabilities towards the formation of inhomogeneities in the $\vec{x}$ or $\theta$ directions.\footnote{We thank Mark Van Raamsdonk for suggesting such a possibility in a discussion on the hard wall model.} In addition to the mere existence of scattering solutions, we have found that the pressure components exhibit clear amplitude modulation, and we have explained this as due to a near-resonance between a scalar shell bouncing frequency and a metric normal mode. It would be interesting to see whether scattering solutions can also be established for more phenomenological holographic models, and if so, whether they exhibit similar features. 

\section{Acknowledgements}
We would like to thank Ioannis Papadimitriou, Joris Vanhoof and Hongbao Zhang for useful discussions. This work was supported in part by the Belgian Federal Science Policy Office through the Interuniversity Attraction Pole P7/37, by FWO-Vlaanderen through project G020714N, and by the Vrije Universiteit Brussel through the Strategic Research Program ``High-Energy Physics''.  EJL is supported by a PhD fellowship from the Research Foundation Flanders (FWO); his work was also partially supported by the ERC Advanced Grant ``SyDuGraM", by IISN-Belgium (convention 4.4514.08) and by the ``Communaut\'e Fran\c{c}aise de Belgique" through the ARC program.

\begin{appendices}
\appendix

\section{Equations of motion suitable for numerics}\label{reom}
In this appendix we will present the equations of motion for the ansatz \eqref{ransatz} used for numerics. We stress that here $'$ will mean derivative with respect to the $r$ coordinate, which is different from the notation in the main text. Also, the $B$ and $\Phi$ below are not the same as in the main text. Defining 
\begin{equation}
\begin{array}{ccc}
P=\dot{b}\frac{f}{h}, &\hspace{20pt} \Pi=\dot{\phi}\frac{f}{h},\\
\Phi=\phi', &\hspace{20pt} B=b',\\
\end{array}
\end{equation}
the equations of motion following from \eqref{ransatz} are
\begin{align}
\frac{h'}{h}=&\frac{r^2}{s^{2d-2}(\frac{gr^2}{s^{2d-2}})'}\Big(\frac{2(d-1)(d-2)}{d}P^2+\frac{4}{d}\Pi^2+g\Phi^2+\\\nonumber
&+\frac{(d-1)(d-2)}{2}gB^2+\frac{2(d-2)(d-1)gB s'}{s}\Big)+(d-2)B-\frac{f'}{f},\label{heq}\\
\dot{P}=&\frac{d}{4(d-1)r}\left(\frac{\left(ge^{(d-1)b}r^2\right)'he^{-(d-1)b}}{fs^{d-1}r}\right)'s^{d-1},\\
\dot{f}=&\frac{(d-2)(d-1)gr^2P h B s + 2(d-2)(d-1) gr^2 P h s'+2gr^2\Pi\Phi hs}{2(\frac{gr^2}{s^{2d-2}})'s^{2d-1}}\nonumber\\
&+\frac{d-2}{2}P h,\\
\frac{h'}{h}=&\frac{8(d-1)(f^2-1)r^2}{(r^2gs^{-2(d-1)})'s^{2d}}+\frac{f'}{f},\label{constraint}\\
\dot{B}=&\left(\frac{P h}{f}\right)',\\
\dot{\Pi}=&\frac{d}{4r}\left(\frac{hg\Phi r}{fs^{d-1}}\right)'s^{d-1},\\
\dot{\Phi}=&(\frac{\Pi h}{f} )',
\end{align}
where $s(r)=z_0(1-r^2)$ and $g(r)=(1-(1-r^2)^d)/(z_0^2r^2d)$.

\section{Normal modes}\label{normalmodes}
To get some analytic understanding of the dynamics of the metric, we will look for normal modes of the metric perturbations. These are solutions of the linearized equations of motion (small perturbations around the soliton background) that can be written as a product of a radial function and a harmonically oscillating function of time. The normal mode spectrum is expected to be discrete in confined geometries, and we saw in Section~\ref{sec:long} that these normal mode frequencies explain the amplitude modulation of the pressure anisotropy.

To find such solutions, we assume that $b=\mathcal{O}(\mu)$, for some small parameter $\mu$, and that the scalar field vanishes. We then solve the equations to linear order in $\mu$. So letting $f=1+\mu f_1+\ldots$, $h=1+\mu h_1+\ldots$, $B=1+\mu B_1+\ldots$ and $P=1+\mu P_1+\ldots$ we obtain
\begin{align}
\dot{f}_1&=z^{1-2d} \frac{(d-2)(d-1)}{(G z^{-2(d-1)})'}P_1G+\frac{d-2}{2}P_1,\\
\dot{P}_1&=\frac{1}{d-1}(G'(h_1-f_1)z^{-(d-1)}+(d-1)B_1Gz^{-(d-1)})'z^{d-1},\\
h_1'&=(d-2)B_1-f_1'+z^{1-2d}\frac{2(d-2)(d-1)GB_1}{(Gz^{-2(d-1)})'},\\
\dot{B}_1&=P_1',\\
h_1'&=\frac{4d(d-1)f_1}{z^{2d}(Gz^{-2(d-1)})'}+f_1',
\end{align}
where $G(z)=1-z^d/z_0^d$. To look for normal modes, we make the ansatz $P_1=Q(z)\cos \omega t$. This implies that $B_1=Q'(z) \sin \omega t /\omega$, $f_1=F(z)\sin\omega t$ and $h_1=H(z)\sin\omega t$, and the functions $Q$, $F$ and $H$ satisfy the ordinary differential equations
\begin{align}
\omega F &= z^{1-2d} \frac{(d-2)(d-1)}{(G z^{-2(d-1)})'}QG+\frac{d-2}{2}Q,\label{Feq}\\
-\omega Q&=\frac{1}{d-1}(G'(H-F)z^{-(d-1)}+\frac{(d-1)Q'G}{\omega }z^{-(d-1)})'z^{d-1},\label{Qeq}\\
H'&=(d-2)\frac{Q'}{\omega}-F'+z^{1-2d}\frac{2(d-2)(d-1)GQ'}{\omega(Gz^{-2(d-1)})'},\\
H'&=\frac{4d(d-1)F}{z^{2d}(Gz^{-2(d-1)})'}+F'.\label{FpHp}
\end{align}

Actually it is possible to extract from these equations a single ordinary differential equation for $Q$. Since $(G'(H-F)z^{-(d-1)})'=G'z^{-(d-1)}(H'-F')$, we can eliminate $H'-F'$ from equation \eqref{Qeq} by using equation \eqref{FpHp}, and then use \eqref{Feq} to eliminate the remaining $F$ such that we end up with
\begin{equation}
-\omega^2 Q=G'\frac{4d}{z^{2d}(Gz^{-2(d-1)})'}\left(z^{1-2d} \frac{(d-2)(d-1)}{(G z^{-2(d-1)})'}QG+\frac{d-2}{2}Q\right)+\left(\frac{Q'G}{z^{d-1}}\right)'z^{d-1},\label{Qeqbis}
\end{equation}
which is a second order ordinary differential equation for $Q$. Demanding regularity in the IR, the only free parameter is $\omega$ (since we can set $Q(z_0)=1$ by an overall rescaling). Then demanding that $Q$ should be normalizable at the boundary (equivalent to $Q(0)=0$), gives us a discrete set of allowed frequencies $\omega$. These are the frequencies of the normal modes, and can be seen in Table \ref{freqtable}.
\begin{table}[H]
\begin{center}
\begin{tabular}{ |c|c|c|c|c| } 
 \hline
 $d$ & 3 & 4 & 5 & 6\\ 
 \hline
 $z_0\omega_0$ & 2.14853 &2.33587  &2.52274&2.70203\\ 
 $z_0\omega_1$ & 4.790 & 5.517 &6.200&6.854\\ 
 $z_0\omega_2$ & 7.116 & 8.069 &8.925&9.719\\ 
 \hline
\end{tabular}
\end{center}
\caption{\label{freqtable} The lowest normal mode frequencies for metric perturbations in various dimensions. The normal mode frequencies are inversely proportional to the confinement scale $z_0$.}
\end{table}

\subsection{Normal modes for the scalar field}
Although not relevant for this work, it is interesting to note that the normal modes of the scalar field satisfy \eqref{Qeq} but with the first term in the RHS, which is proportional to $Q$, removed:
\begin{equation}
-\omega^2 Q=\left(\frac{Q'G}{z^{d-1}}\right)'z^{d-1},\label{Qeqscalar}
\end{equation}
if we assume that $\phi=Q\cos\omega t$. Given that the omitted term is proportional to $Q$, and therefore combines with the LHS, the normal modes of the scalar and the metric fluctuations can be expected to approach each other for large $\omega$. In addition, we have observed numerically that the spectrum becomes linear for large $\omega$. Global AdS$_3$ is actually identical to the AdS$_3$ soliton, which leads us to expect that \eqref{Qeqscalar} must give a linear spectrum for $d=2$. Indeed, in this case if we redefine $Q=z^2q$ and $z^2/z_0^2=(x+1)/2$, we obtain the equation
\begin{equation}
(1-x^2)q''+(1-3x)q'+(\frac{\omega^2}{4}-1)q=0,
\end{equation}
which is solved by the Jacobi polynomials $q(x)=P_n^{(\alpha,\beta)}(x)$ with $\alpha=0,\beta=1$ and with $\omega=2(n+1)$ for $n=0,1,2,\ldots$.

\section{Asymptotic expansion for $d=3$}\label{asympt3}
Here we provide the complete asymptotic expansion for $d=3$, including the time window when the source is turned on. We will thus assume a source $J_b$ for the function $b$ and a source $J_\phi$ for the scalar field $\phi$. The asymptotic expansions for the various functions, following from the equations of motion, are then
\begin{equation}
f(z,t)=1+\frac{1}{8}(\dot{J}_b^2+\dot{J}_\phi^2)z^2+\frac{E}{4}z^3+O(z^4),
\end{equation}
\begin{equation}
h(z,t)=1-\frac{1}{4}(\dot{J}_b^2+\dot{J}_\phi^2)z^2-\frac{E}{4}z^3+O(z^4),
\end{equation}
\begin{equation}
b=J_b-\frac{1}{2}\ddot{J}_bz^2+b_3z^3+O(z^4),
\end{equation}
\begin{equation}
\phi=J_\phi-\frac{1}{2}\ddot{J}_\phi z^2+\phi_3z^3+O(z^4).
\end{equation}
We also obtain the Ward identity
\begin{equation}
3(2b_3-\frac{1}{z_0^3})\dot{J}_b+6\dot{J}_\phi \phi_3+2\dot{E}=0.
\end{equation}
When going to Fefferman-Graham gauge as in Section~\ref{boundaryexp}, the intermediate $z^2$ terms will not affect the $z^3$ terms. Moreover, since in even dimensional AdS spaces the counterterms do not affect the field theory observables, the boundary observables are given by the same formulas \eqref{Ttt} and \eqref{vev}, even when the sources are turned on. The Ward identity then takes the form
\begin{equation}
3\left(\langle T_{\theta\theta}\rangle-\langle T_{xx}\rangle\right)\dot{J}_b+6\langle \mathcal{O}\rangle\dot{J}_\phi+2\langle T_{tt} \rangle=0.
\end{equation}

\end{appendices}

\end{document}